\documentclass[preprint,eqsecnum,aps]{revtex4}     
\usepackage{graphics}
\def\beq{\begin{equation}}
\def\eeq{\end{equation}}
\def\bea{\begin{eqnarray}}
\def\eea{\end{eqnarray}}
\def\nnu{\nonumber}
\def\tst{\textstyle}

\def\fno#1{Fig.~\ref{#1}}
\def\ftwo#1#2{Figs.~\ref{#1} and \ref{#2}}

\def\eno#1{Eq.~(\ref{#1})}
\def\Eno#1{Equation (\ref{#1})}
\def\etwo#1#2{Eqs.~(\ref{#1}) and (\ref{#2})}

\def\Sno#1{Sec.~\ref{#1}}

\def\by{\over}
\def\gtwid{\mathrel{\raise.3ex\hbox{$>$\kern-.75em\lower1ex\hbox{$\sim$}}}}
\def\ltwid{\mathrel{\raise.3ex\hbox{$<$\kern-.75em\lower1ex\hbox{$\sim$}}}}

\def\al{\alpha}

\def\dta{\delta}
\def\eps{\epsilon}

\def\kap{\kappa}

\def\sig{\sigma}

\def\Gam{\Gamma}
\def\Dta{\Delta}

\def\Tta{\Theta}

\def\apx{\approx}
\def\ptl{\partial}

\def\hf{{1\over2}}
\def\tshf{{\tst\hf}}

\def\sumpr{\mathop{{\sum}'}}
\def\inmipi{\int_{-\infty}^{\infty}}

\def\grad{\nabla}

\def\part#1#2{{\ptl#1 \by \ptl#2}}

\def\lf{\left}
\def\rt{\right}

\def\ham{{\cal H}}

\def\avg#1{\langle#1\rangle}

\def\bB{{\bf B}}
\def\bH{{\bf H}}

\def\bM{{\bf M}}
\def\bS{{\bf S}}

\def\nhat{{\bf{\hat n}}}

\def\ity{{\it y\ }}
\def\itz{{\it z\ }}
\def\itxy{{\it xy\ }}
\def\iti{{\it i\ }}

\def\tus{t_{\rm us}}
\def\Fe8{Fe$_8$}
\def\Mn12{Mn$_{12}$}

\def\up{\uparrow}
\def\dn{\downarrow}

\def\pin#1{p_{{\rm in},#1}}
\def\pout#1{p_{{\rm out},#1}}
\def\pinb{p_{{\rm in}}}
\def\poutb{p_{{\rm out}}}
\def\pfl{p_{\rm flip}}
\def\rb{{\bar r}}
\def\sr{{\cal S}_r}
\def\srup{{\cal S}_{r\up}}
\def\srdn{{\cal S}_{r\dn}}
\def\srb{{\cal S}_{\rb}}
\def\srbup{{\cal S}_{\rb\up}}
\def\srbdn{{\cal S}_{\rb\dn}}
\def\sigt{\tilde\sig}
\def\sumprj{\sumpr_{\!\!j}}

\begin{document}


\title{Low-Temperature Magnetization Relaxation in Magnetic Molecular Solids}
\author{Avinash Vijayaraghavan}
\author{Anupam Garg}
\email[e-mail address: ]{agarg@northwestern.edu}
\affiliation{Department of Physics and Astronomy, Northwestern University,
Evanston, Illinois 60208}

\date{\today}

\begin{abstract}
The low temperature relaxation of the magnetization in molecular magnetic solids
such as \Fe8 is studied using Monte Carlo simulations. A set of rate equations
is then developed to understand the simulations, and the results are compared.
The simulations show that the magnetization of an intially saturated sample
deviates as a square-root in time at short times, as observed experimentally,
and this law is derived from the rate equations analytically.
\end{abstract}

\maketitle

\newpage
\section{Introduction}
\label{intro}

The low temperature relaxation of the magnetization of magnetic
molecular solids such as \Fe8 has proven difficult
to understand ever since the earliest experimental studies~\cite{san97,ohm,ww4}.
The time dependence of this relaxation is highly non-exponential, and
fits to forms such as stretched exponentials have provided no
insight even when the fits seem to be good. A second puzzling
feature is that
for short times, the relaxation is observed to follow a
square-root behavior with time in a large number of protocols:
demagnetization, magnetization, and hole-digging in which the
magnetic field is abruptly changed after the magnetization has
been allowed to come to an equilibrium or quasi-equilibrium state
in response to a previous value of the applied
magnetic field. A good review of the subject is given by
Gatteschi, Sessoli, and Villain~\cite{gsvbook}. These authors give
many more references to experimental
studies \cite{san97,ohm,ww4,tcb,ww5,ww3,tb}, theoretical analyses \cite{pro98ab},
and Monte Carlo simulations \cite{Cuc99,Fer0304}.

The fundamental microscopic mechanism by which the spin of an
individual molecule changes at low temperatures (say below 50 mK)
is incoherent tunneling
between the lowest energy states. In both \Fe8 and \Mn12, the
anisotropy of the molecule is of the Ising type, and the lowest
energy states have Zeeman quantum numbers $m = \pm S$, where $S$
is the spin of the molecule. The tunnel splitting between these
states is of order 100~Hz (in frequency units) for \Fe8 and
unobservably small for \Mn12. It must be stressed that in the
solid, the tunneling is not of the coherent flip-flop type seen
in the NH$_3$ molecule, and previous authors have examined
various decoherence processes by which the tunneling
dynamics of a single molecule change from coherent to incoherent \cite{pro96,avag09}.
This is not enough to explain the observed non-exponential time
behavior, for if there were a single characteristic time
scale for relaxation of a single molecule, and all molecules
relaxed independently, the magnetization would relax essentially
exponentially in time with the same time scale as for one molecule.
Thus, the non-exponential time behavior is a 
strong indicator that the
molecules in the solid do not relax independently of each other.
The biggest and most obvious coupling between molecules is the
dipole-dipole interaction, and while this has been considered by many
previous authors \cite{ohm,pro98ab,gsvbook} a complete theory is still lacking.
In particular, the $\sqrt{t}$ form has been previously explained by
Prokofeev and Stamp~\cite{pro98ab}, but
as noted by Gatteschi, Sessoli, and Villain~\cite{gsvbook}
it is unclear if it applies to all situations. These latter authors also give a
heuristic argument for the $\sqrt{t}$ law for the particular case of the
demagnetization problem. We comment further or this below.

In this paper, we report on our attempt to
solve this problem. Our first approach is Monte Carlo simulation.
In this we follow in the footsteps of Refs.~\cite{Cuc99,Fer0304}, and many
aspects of our simulation and the results are very similar to those found
by these authors. We then try to
understand our Monte Carlo results by developing a set of rate
equations. These rate equations entail the distribution of dipole
fields at the molecular sites. For the specific problem of
demagnetization, we can construct an approximate model for this
distribution, which then enables us to solve the rate equations
numerically. We find that the solution to the rate equations
matches the Monte Carlo results quite closely. Furthermore,
we can show analytically that that the
solution obeys a square-root behavior with time at
short times. We emphasize that as in Ref.~\cite{gsvbook} we have only studied the
demagnetization problem. Further, the scaling behavior that we find for
ancillary quantities also agrees entirely with Ref.~\cite{gsvbook}.
Thus, we can claim no priority for this result.

We also emphasize that our model for the dipole field
distributions and the rate equations requires no further
ingredients or fitting parameters beyond those involved in
specifying the Monte
Carlo process. In this paper we only look at the problem of
demagnetization of a spherical sample with a cubic lattice in
order to minimize the complications from demagnetizing fields,
and focus on the shape independent aspects of the problem,
but we believe that our rate equation approach offers a method
to attack a much wider class of problems, and in the future we
hope to study other experimental protocols, sample shapes, and
lattice types.

The plan of the paper is as follows. In \Sno{phys_mod} we describe
the basic physical model underlying the relaxation \cite{pro98ab,avag09}.
We then describe our Monte Carlo simulations and results in \Sno{montecarlo}.
The theory for the rate equations and the bias distribution are developed
in Secs.~\ref{theory} and \ref{bias_dist}. Finally, in \Sno{root_t} we
present our analytical solution to the rate equations, and the
$\sqrt{t}$ law.

\section{Physical Model for Relaxation}
\label{phys_mod}
As shown in Ref.~\cite{avag09}, the fundamental process that
governs the dynamical behaviour of the spins is as follows. In a
short time interval $dt$, the spin of the $i$th molecule flips
from $m=-S$ to $m=S$, or $m=S$ to $m=-S$, with a probability
\beq
p_{{\rm{flip}},i} = \Gam_i dt,
\eeq
with
\beq
\Gam_i \equiv \Gam(E_i)
   = {\sqrt{2\pi} \by 4} {\Dta^2 \by W}
          \exp -\left({E_i^2\by 2W^2} \right).  \label{p_flip}
\eeq
Here, $\Dta$ is defined via the statement that $i\Dta/2$ is the quantum
mechanical amplitude per unit time for a spin to tunnel between
the $m = \pm S$ states, $W \simeq 10 E_{dn}$, where $E_{dn}$ is the 
energy of dipole-dipole interaction between the molecular electronic
spin and the nuclear spins of nearby nonmagnetic atoms such as N and H which
are always present in the molecules studied, and $E_i$ is the energy of the
$m = S$ state relative to the $m=-S$ state due to the net magnetic field
seen by the $i$th molecule. We shall refer to $E_i$ as the bias on site
$i$~\cite{bias}.
For \Fe8, $\Dta \sim 10^{-8}$~K, $E_{dn} \sim 1$~mK, and $E_i \sim 0.1$~K in
temperature units. (We shall set $\hbar$ and $k_B$ to unity in all working
formulas, so temperature, energy, and frequency all have the same units.)

The dominant feature in \eno{p_flip} is the exponential suppression
of the flip rate with the square of the bias energy $E_i$, and a large part of this
energy arises from the dipolar field of the other molecular spins
in the solid, which can be estimated to be of order $100$~Oe for
near neighbour spins, leading to the energy scale $0.1$~K quoted above.
More explicitly, the dipolar part of $E_i$ is given by
\bea
E_{i,\rm{dip}} &=& \sum_{j\ne i} K_{ij} \sig_j, \label{Ebias} \\
K_{ij} &=& 2 {E_{dm} a^3 \by r^3_{ij}}
                      \lf(1-3{z^2_{ij} \by r^2_{ij}} \rt). \label{Kij}
\eea
Here, $E_{dm}$ is the energy scale of interaction for near neighbours,
$a$ is the near-neighbour distance, $r_{ij}$ is the distance between
spins $i$ and $j$, $z_{ij}$ is the projection of the corresponding displacement onto
the {\it z\/}-axis, the easy axis of the spins. Finally, $\sig_i$ is an
Ising spin variable such that $\sig_i = \pm 1$ when the true spin on site
$i$ is $\pm S$.

Since the dipole field is long ranged, and $E_{dm} \gg E_{dn}$, the
flip of the $i$th spin changes the bias field on a large
number of neighbouring spins, and thus changes the flip probability for
those spins significantly. The relaxation of the magnetization of the
entire solid is therefore a complex coupled process in which every
individual spin essentially waits until it experiences a bias field
less than $W$ in magnitude, and then flips with a probability per unit
time equal to approximately $\Dta^2/W$. The flip of this spin changes the
bias field at many other molecules, and if one of them then happens to
have a near-zero bias field, it flips, leading to the possibility of
flips at yet more molecules. Ref.~\cite{gsvbook} refers to this scenario as a
long-range Glauber model.
\section{Monte Carlo Simulation}
\label{montecarlo}
\subsection{Simulation protocol}
\label{protocol}
As explained in \Sno{intro}, in this paper we only report on simulations
on spherical samples of $N$ spins on a cubic lattice in order to eliminate
the effects of inhomogeneous demagnetizing fields. In addition, we only
consider the demagnetization process. Thus, the spin $\sig_i$ is initialized
to the value +1 at every site. Starting from this configuration, we simulate
the time evolution of the sample (as described below) for between 60 and 500
runs, and then average the total magnetization of the entire sample over
these runs. We have performed simulations for two sample sizes, with
$N = 9,171$ and 82,519.

The initial spin polarization creates an almost delta-function-like
distribution of bias fields centered at zero field, exactly as expected
theoretically. We see small deviations from a perfectly uniform distribution
due to the finite size of the sphere.

The evolution of the system from time step $t$ to the next time step
$t + dt$ is carried out using the following protocol. At time $t$,
the bias energy $E_i$ is computed at every site using \eno{Ebias}. All
spins are then flipped or not flipped using the flipping protocol
described below. We are now at time $t + dt$. The bias fields are
recomputed at all sites, and the process is repeated.

The flipping protocol we employ entails a slightly modified flip
probability
\beq
p_{{\rm{flip}},i} = {\Dta_2^2 \by 4W}\, \Tta(W - |E_i|)\,dt,
     \label{p_flip2}
\eeq
instead of the original form (\ref{p_flip}). Here,
$\Tta(\cdot)$ is the Heavyside function equal to unity for positive
argument and to zero for negative argument. In other words, a spin flips
only if the bias field on it is less than $W$ in magnitude. This modification
is not material to the physics, and it reduces the run time of the
simulations. We refer to the spins in the window $|E_i| < W$ as
{\it reversible\/}. We have also used \eno{p_flip} in a few cases, and
not found any
significant differences in the results. Further, $\Dta_2 = \sqrt{\pi} \Dta$,
and the prefactor in \eno{p_flip2} is chosen to ensure that the integral
$\inmipi p_{{\rm flip}}(E) dE$ is unchanged. In this way, the total
magnetization that flips in a large subvolume containing many spins is
unaffected. For future use we define
\beq
\Gam_0 = {\Dta_2^2 \by 4W}.
\eeq

An important consideration arises with regard to the values of $E_{dm}$,
$\Dta$, and $W$ to be used in the simulation. We know that the ratio of
these quantities for real \Fe8 is $E_{dm}/\Dta \sim 10^7$, and
$E_{dm}/W \sim 10$. Due to the long ranged nature of the dipole field,
when a spin flips, it has the potential to bring $\sim 10 E_{dm}/W$
spins into the {\it reversibility\/} region $|E_i| < W$. We refer to
this as the {\it influence sphere\/} of the spin. To overcome finite
size effects, we must make sure that our simulation includes a large
number of influence spheres. Secondly, the rate at which a spin flips,
even if it is within the reversibility window, is governed by $\Dta$,
and our simulation would be much too slow if we used the actual
value of $\Dta/E_{dm}$. We have therefore chosen different values for
these quantities while still ensuring the physically important
restriction $E_{dm} \gg W \gg \Dta$. Specifically, we take
$\Dta_2 = 2.0$, $E_{dm} = 50\Dta_2$, and vary $W$ over a range of
values between $\Dta_2$ and $E_{dm}$.

The next consideration is over the choice of the time step $dt$. We
set $dt = 0.01 E_{dm}/\Dta_2^2$, and hence independent of $W$. This is
done in order to remain true to the idea that the flip probability for
a reversible spin should depend on $W$ only through the rate
$\Gam_0$, and not $dt$. With our choice of $dt$ this probability is
\bea
p_{\rm flip} &=& {\Dta_2^2 \by 4W} dt \nnu \\
             &=& 0.01 {\Dta_2^2 \by 4W} {E_{dm} \by \Dta_2^2} \nnu \\
             &=& 0.01 {E_{dm} \by 4W}.
\eea
By choosing $E_{dm}/4W \ltwid 10$, we ensure that the flip probability in
one time step is not too large, which in turn ensures that our
discretization of time is not too coarse, and that the simulation is
sufficiently close to a continuous process. At the same time, $p_{\rm flip}$
is large enough that we do not expend unnecessary time steps in waiting for
the spin configuration to change by a meaningful amount. The time-scale
$\tau = E_{dm}/\Dta_2^2$ demarcates short versus long times, and we shall
study relaxation for $\sim 10^3 \tau$ in some cases, i.e., $\sim 10^5$
time steps. For real \Fe8 we have $\tau \simeq 10^4$ secs.

Some other details of the simulation are as follows. The spherical sample
is built from a cube having an odd number of sites on a simple cubic
lattice with lattice constant `a', and selecting those sites within a
distance $Da/2$ of the origin in order to get a sphere of diameter
$Da$. The two system sizes $N = 9,171$ and $N = 82,519$ correspond to
sphere diameters $D = 27$ and $D = 55$ respectively. The sites are
indexed from 1 to $N$, and their Cartesian coordinates
are stored in one-dimensional arrays. To reduce computer time, at the
start a one-dimensional look-up table is made of the kernel $K_{ij}$ by
converting the triple of distances ($x_{ij}, y_{ij}, z_{ij}$) into a
single unique number using some artificial but easy-to-implement formula
that is invertible, i.e., capable of yielding the
triple ($x_{ij}, y_{ij}, z_{ij}$) from the single number.
\subsection{Quantities Measured}
\label{measured}
The central quantity of interest that is measured in our simulations is
the magnetization,
\beq
m = (N_{\up} - N_{\dn})/N, \label{def_mag}
\eeq
where 
$N_{\up}$ and $N_{\dn}$ are the number of up and down spins. The magnetization
is measured at every time step.

In addition, we also measure at every time step, the bias distribution
$\rho(E)$, defined such that $\rho(E) dE$ is the fraction of spins experiencing
a bias field between $E$ and $E + dE$. The bin width for numerical purposes is
chosen as $W$ itself as this is a sufficiently small number compared to $E_{dm}$.
Secondly, the distribution is measured for biases that satisfy
$|E_i| \le 15 E_{dm}$. In practice, we find that the fraction of sites
that lie outside this range is $O(10^{-2})$.
\subsection{Results of the Simulations}
\label{sim_results}
As mentioned above, we have performed the simulations for different relative
values of $W$ and $E_{dm}$. For a test case, we made the contraphysical choice
$W \gg E_{dm}$. In this case we expect each spin to remain reversible most
of the time, and rarely move out of the reversibility window when neighbouring
spins flip. Each spin should then relax essentially independently of the others,
leading to exponential relaxation of the magnetization with a rate $2\Gam_0$.
This is indeed what is observed, giving us confidence in our numerical code.

The physically interesting simulations are performed for $E_{dm} \gg W$. In
\fno{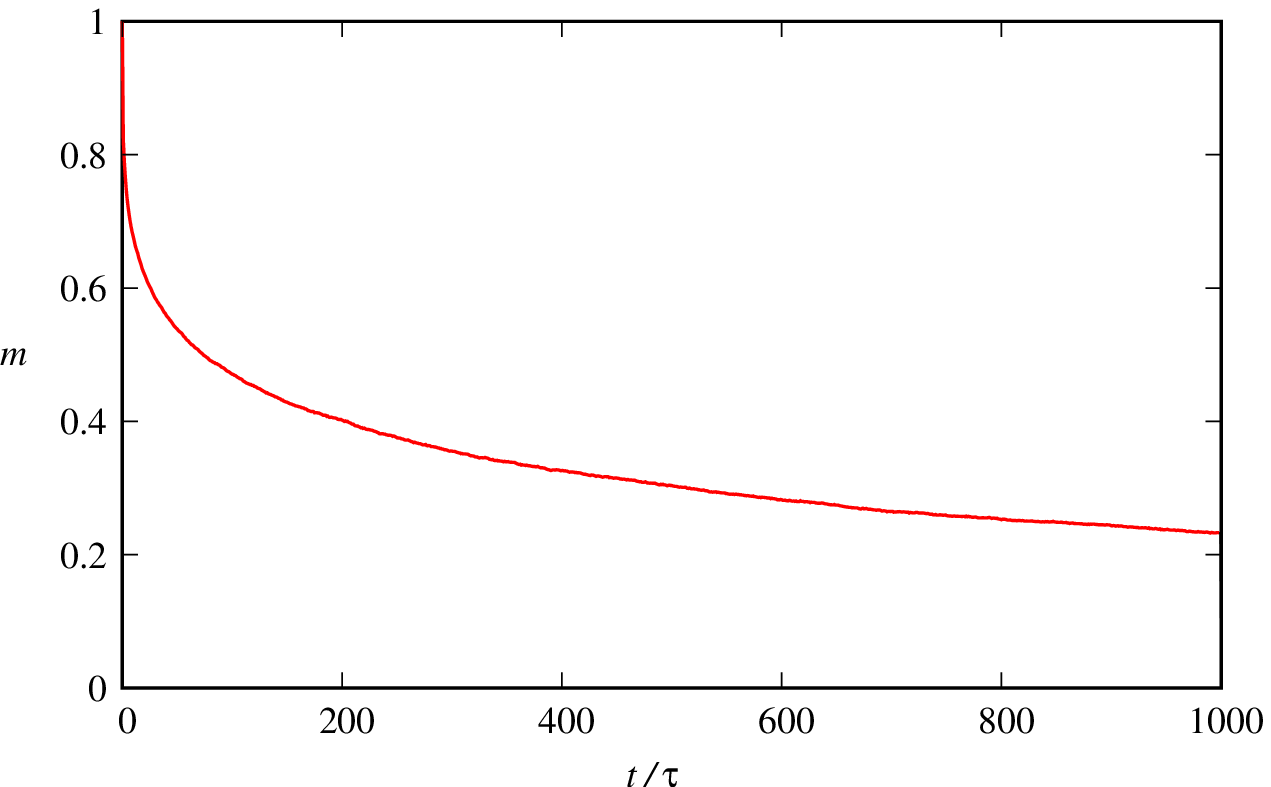}, we show the magnetization versus time for one such simulation
over a time $1000\tau$. It is evident that the decay of $m$ is nonexponential,
and that there is a steep initial drop in $m$ over a time of order $\tau$.
This drop is shown in more detail in \fno{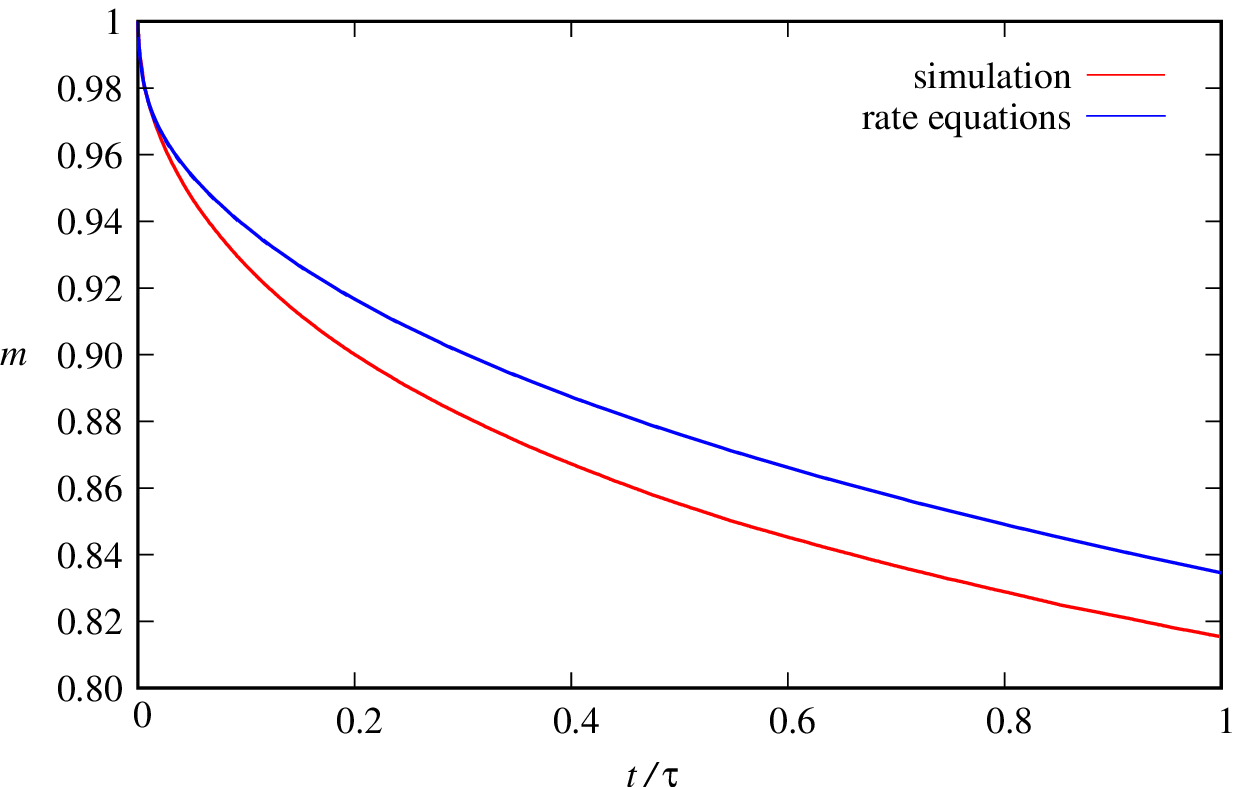}, and is quite well fit by a
square-root form; we discuss this in more detail in \Sno{root_t}. In both these figures,
we have performed an average over 60 runs.
\begin{figure}
\includegraphics{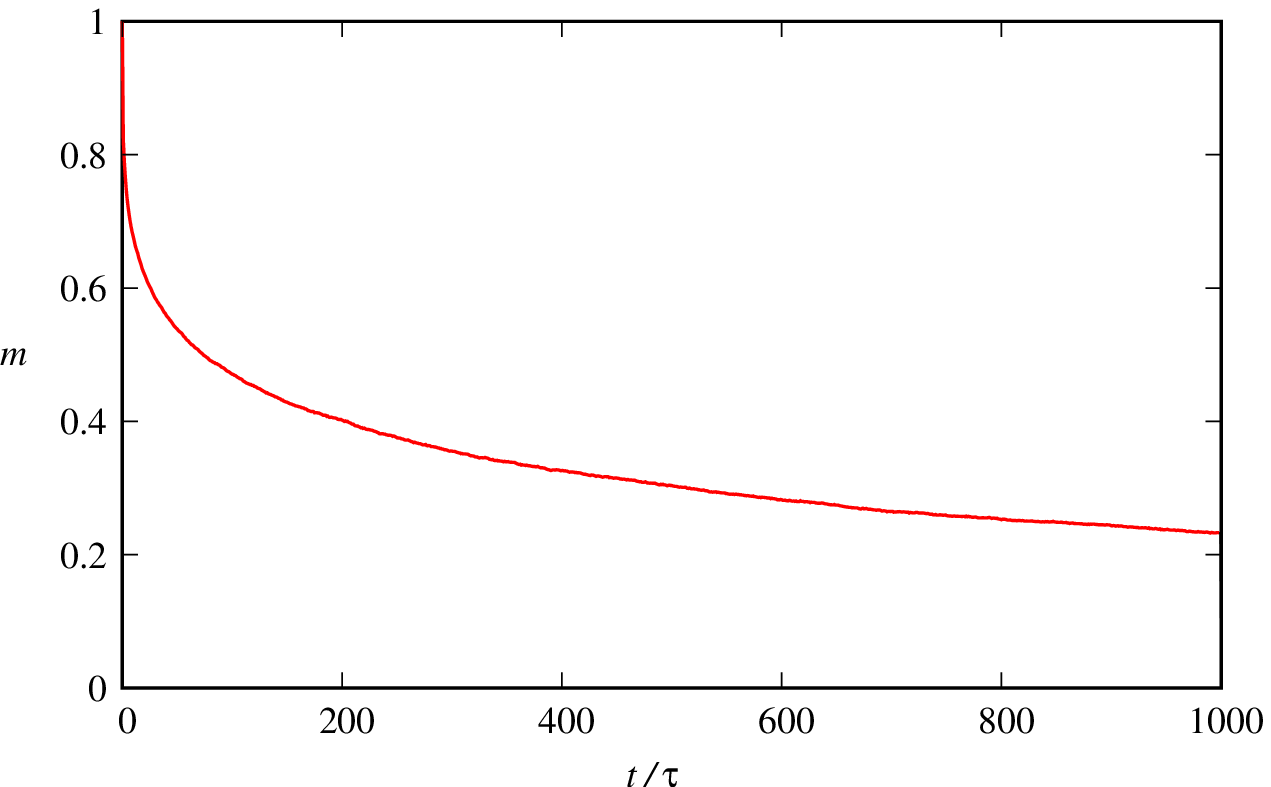}
\caption{\label{m_vs_t_long.eps}
(Color online) Long-time decay of magnetization for the $N=82519$ spin sample, averaged
over 60 runs. The parameter values are $W = 2.5\Dta_2$, and $E_{dm} = 50\Dta_2$. 
} 
\vskip15pt
\includegraphics{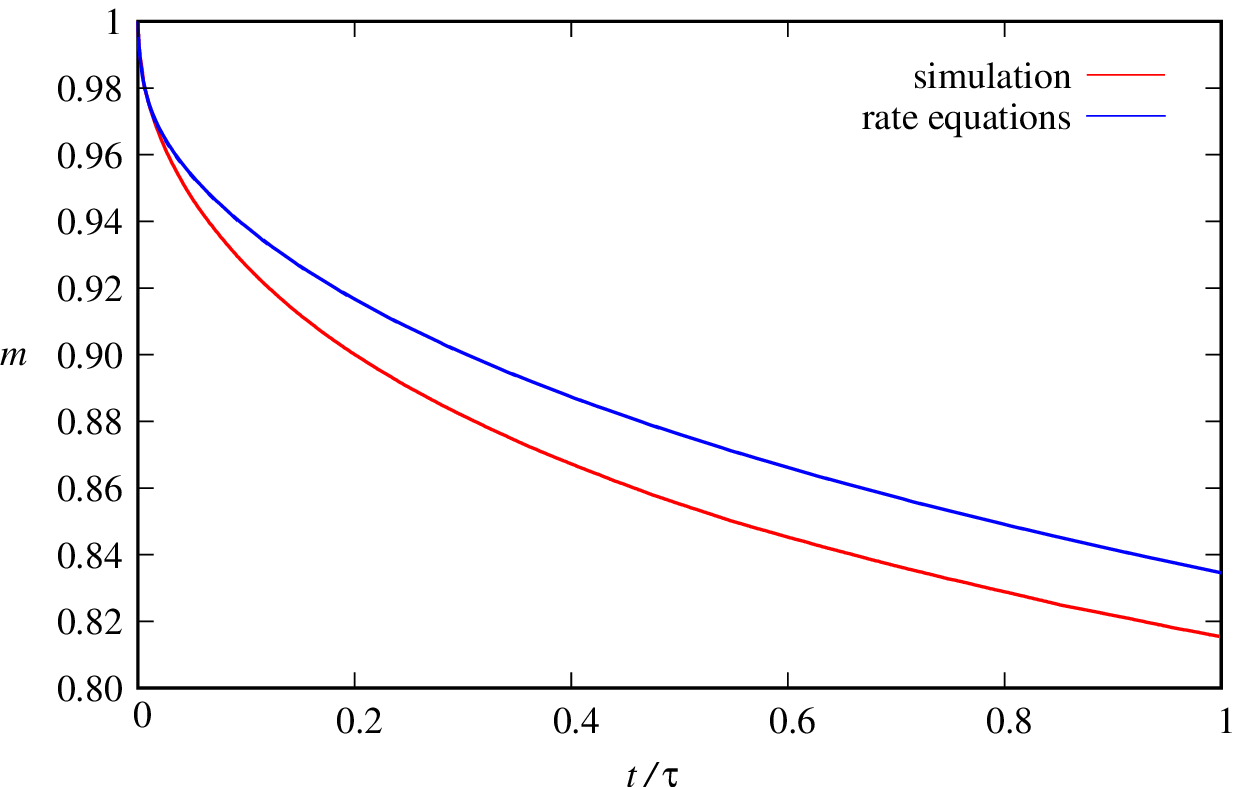}
\caption{\label{m_vs_t_short.eps}
(Color online) Short-time behavior of the magnetization, with the same parameters
as \fno{m_vs_t_long.eps}. Also shown is the result from numerical solution of the
rate equations.
} 
\end{figure}

In \ftwo{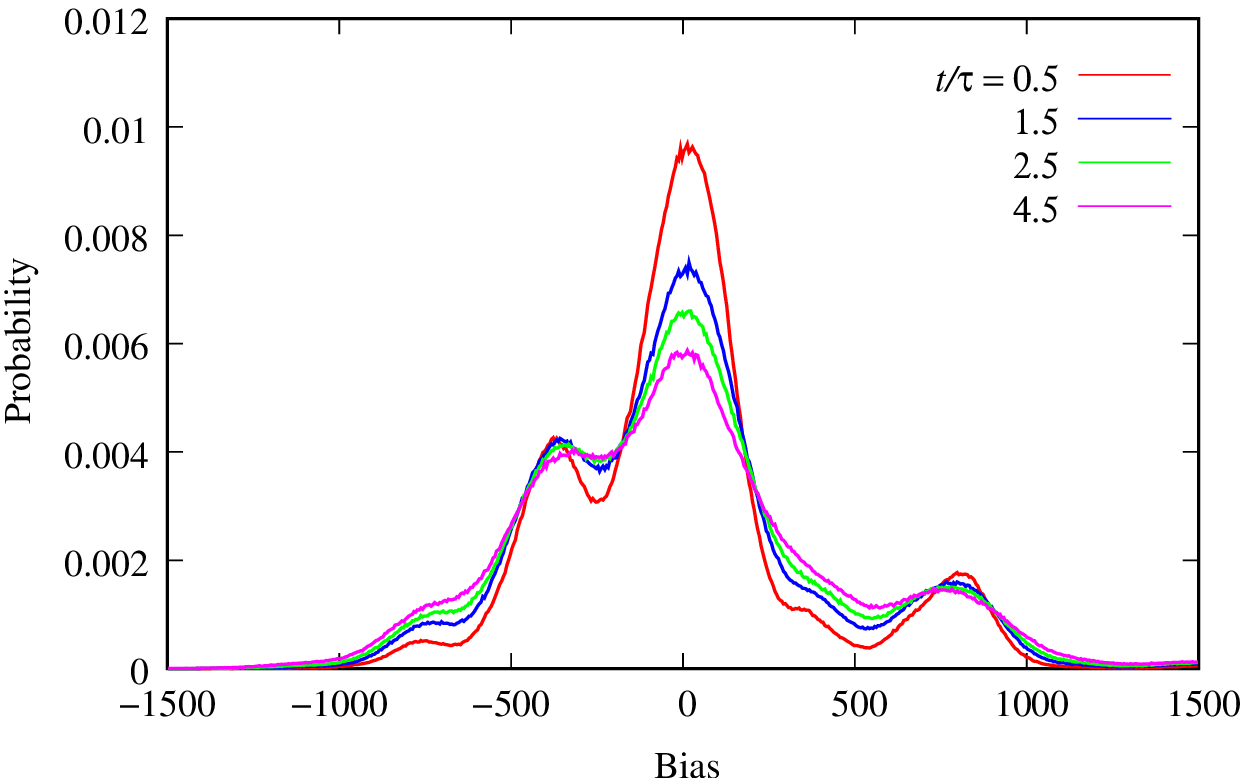}{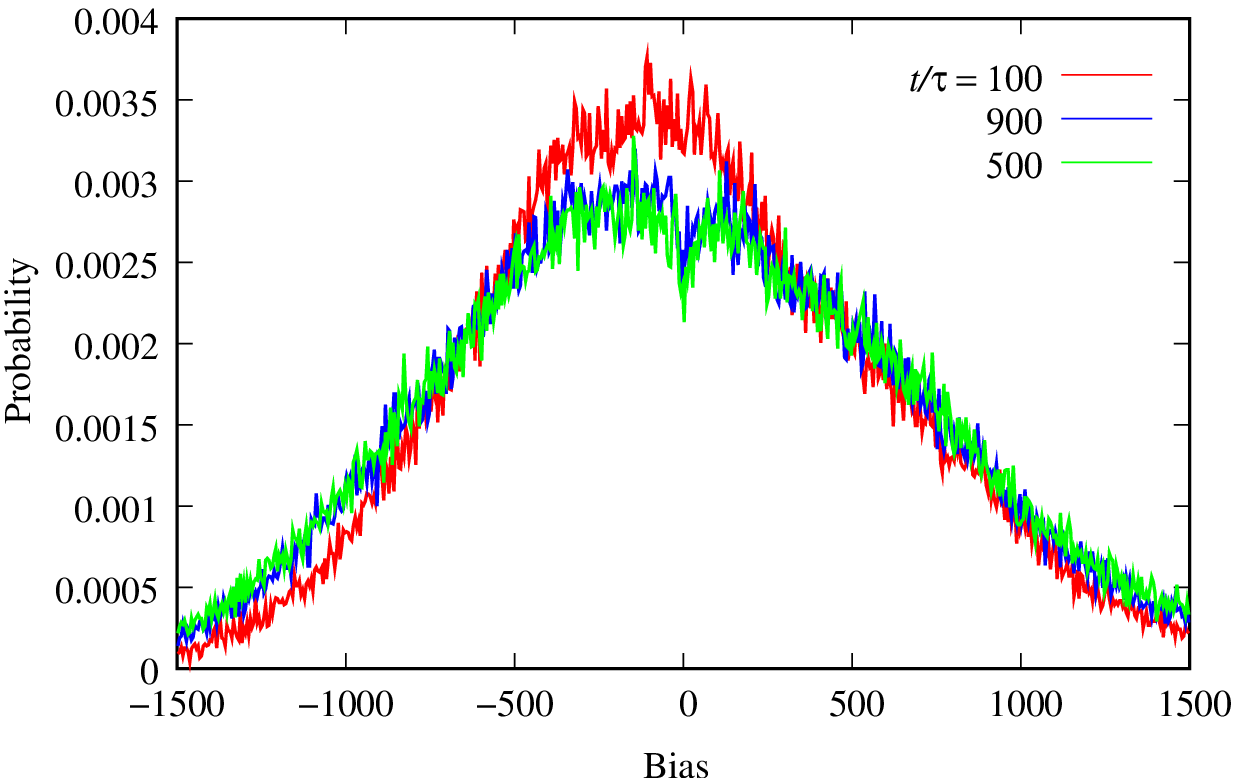} we show the short- and long-time bias
distribution $\rho(E)$ for the same parameters as in Figs.~\ref{m_vs_t_long.eps} and
\ref{m_vs_t_short.eps}. At short times, the distribution is marked by three clear
peaks, as well as a few shoulders, which we shall explain in more detail in
\Sno{bias_dist}. Here we note that the two main peaks other than at the center
are at $-4E_{dm}$ and $8 E_{dm}$. It is also evident that the peaks and shoulders become
less distinct as $t$ increases. Indeed, for $t \ge 100\tau$, they disappear completely, as
shown in \fno{dist_long_82k.eps}. Here we see a new feature developing, namely a hole in the
distribution at $E = 0$, for $t \ge 500\tau$.
\begin{figure}
\includegraphics{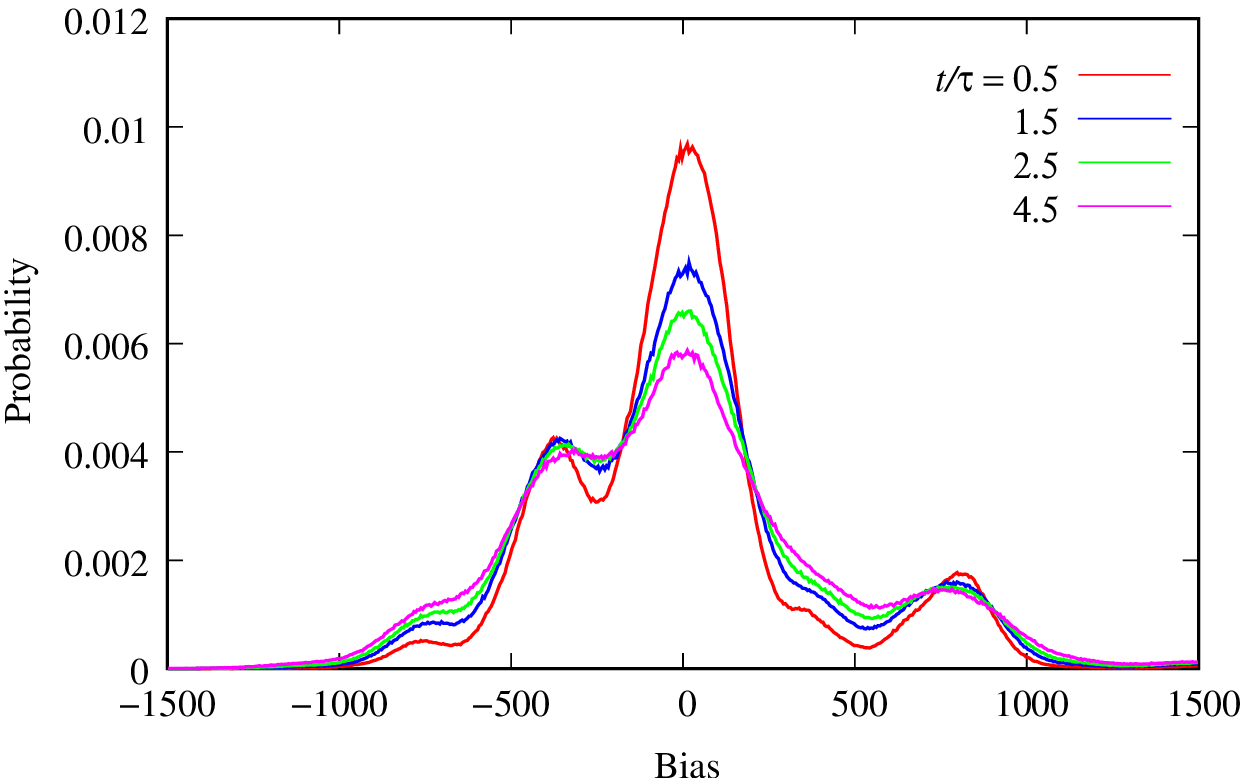}
\caption{\label{dist_short_82k.eps}
(Color online) Histogram of the short-time bias distribution for the $N=82519$ spin sample, averaged over 60
runs, with the same parameters as in Figs.~\ref{m_vs_t_long.eps} and \ref{m_vs_t_short.eps}.
The bin width in the bias is $5.0$.
} 
\vskip15pt
\includegraphics{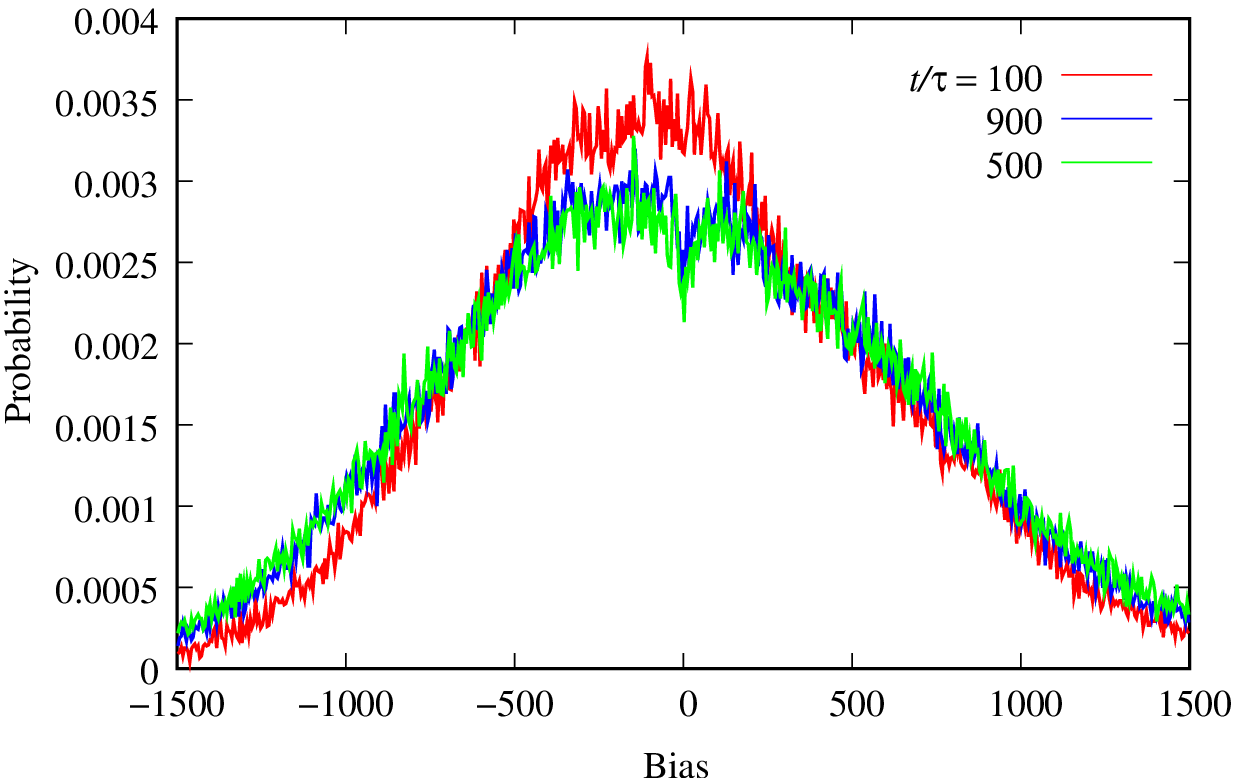}
\caption{\label{dist_long_82k.eps}
(Color online) Same as \fno{dist_short_82k.eps} but for long times, and averaged over 10 runs only. Note the
reduced scale on the \ity axis.
} 
\end{figure}

The bias distributions also provide a good indicator of whether our system size is large
enough and whether the averaging procedure is valid. To this end, we show in
\ftwo{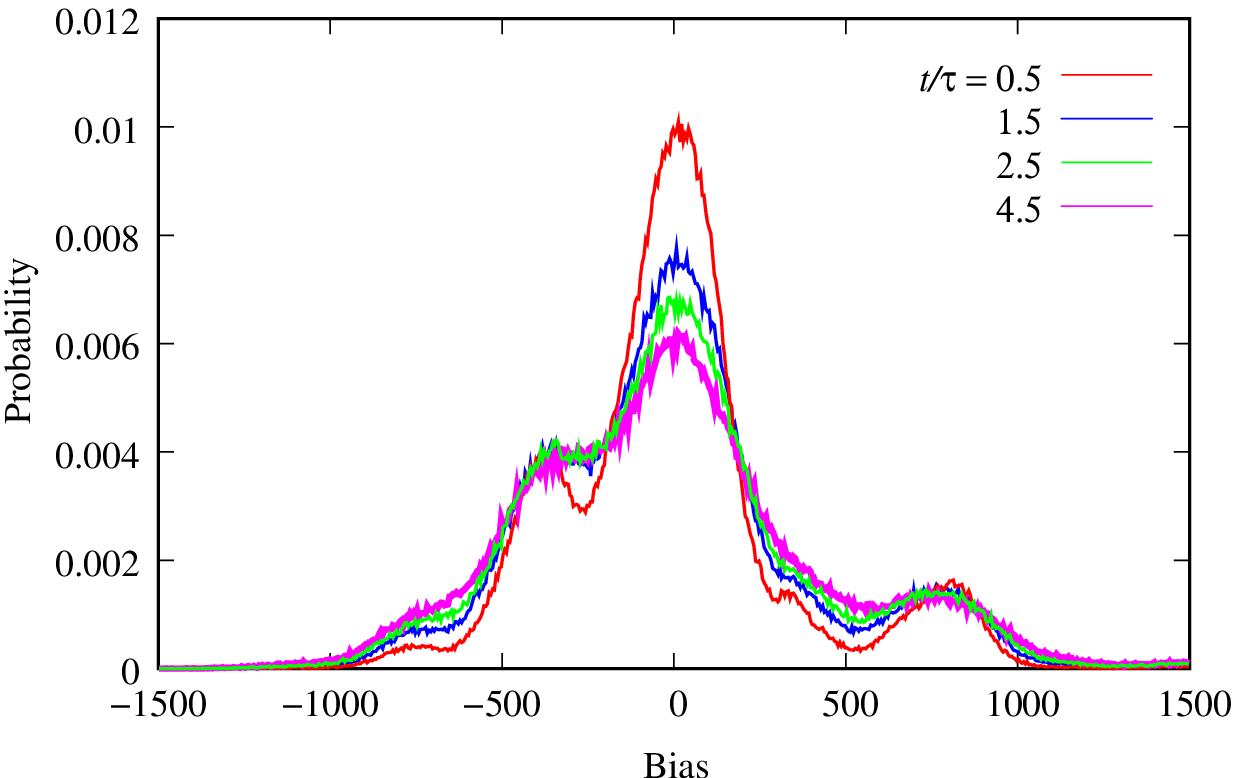}{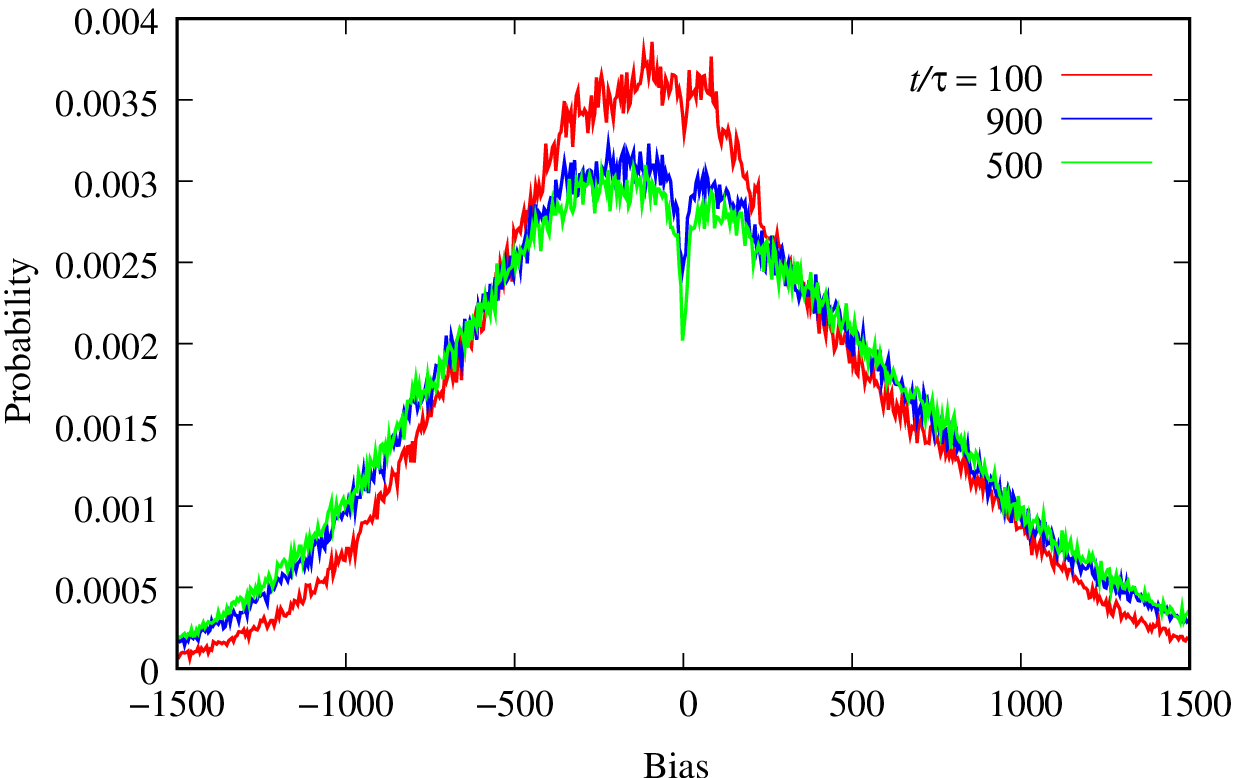} the short- and long-time distributions for the
smaller sample size ($N = 9171$) but all parameters the same as in previous figures. The
two figures are drawn for averages over 60 and 30 runs, respectively. As can be seen, the
statistical scatter is only minimally greater, and the quantitative features---heights
and locations of the peaks at short times, the hole at zero bias at long times---are
identical. Finally, in
\fno{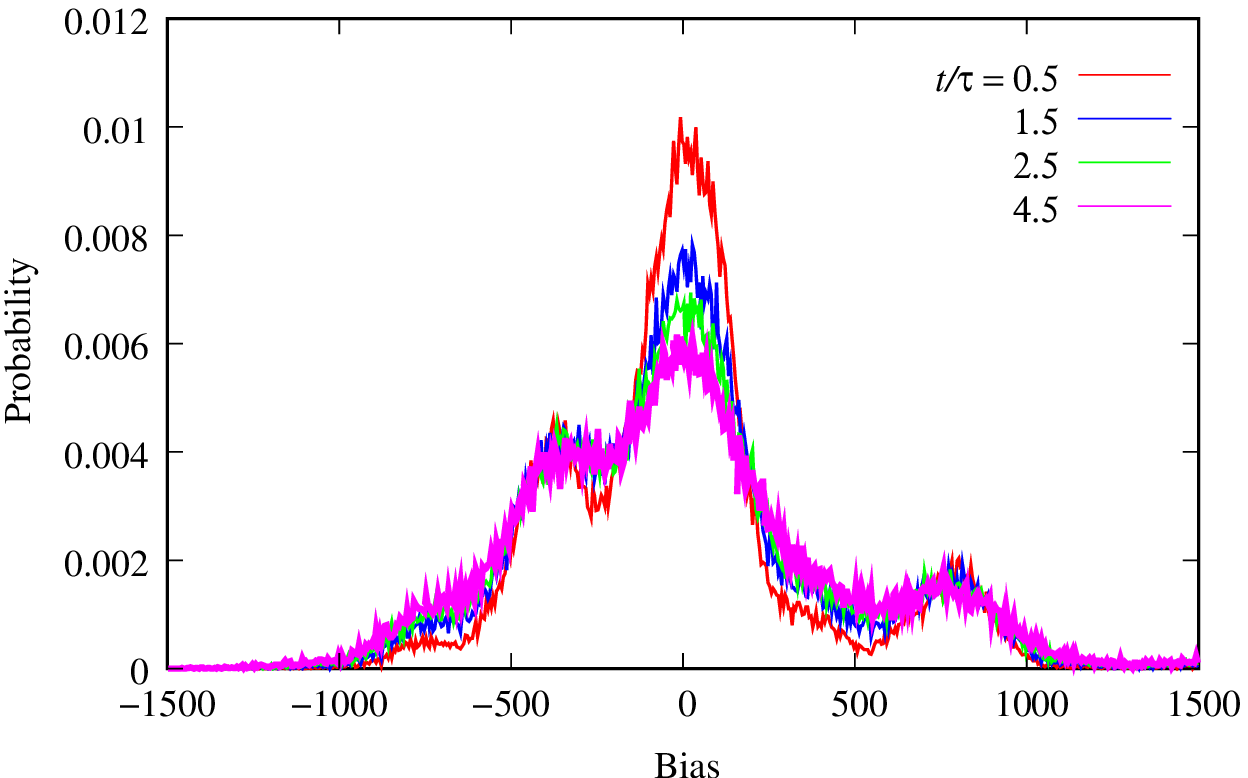}, we show the short-time distribution for a single run of the larger
sample. The features seen in the 60-run average are all clearly present, showing that
questions of self-averaging do not arise in this system.
\begin{figure}
\includegraphics{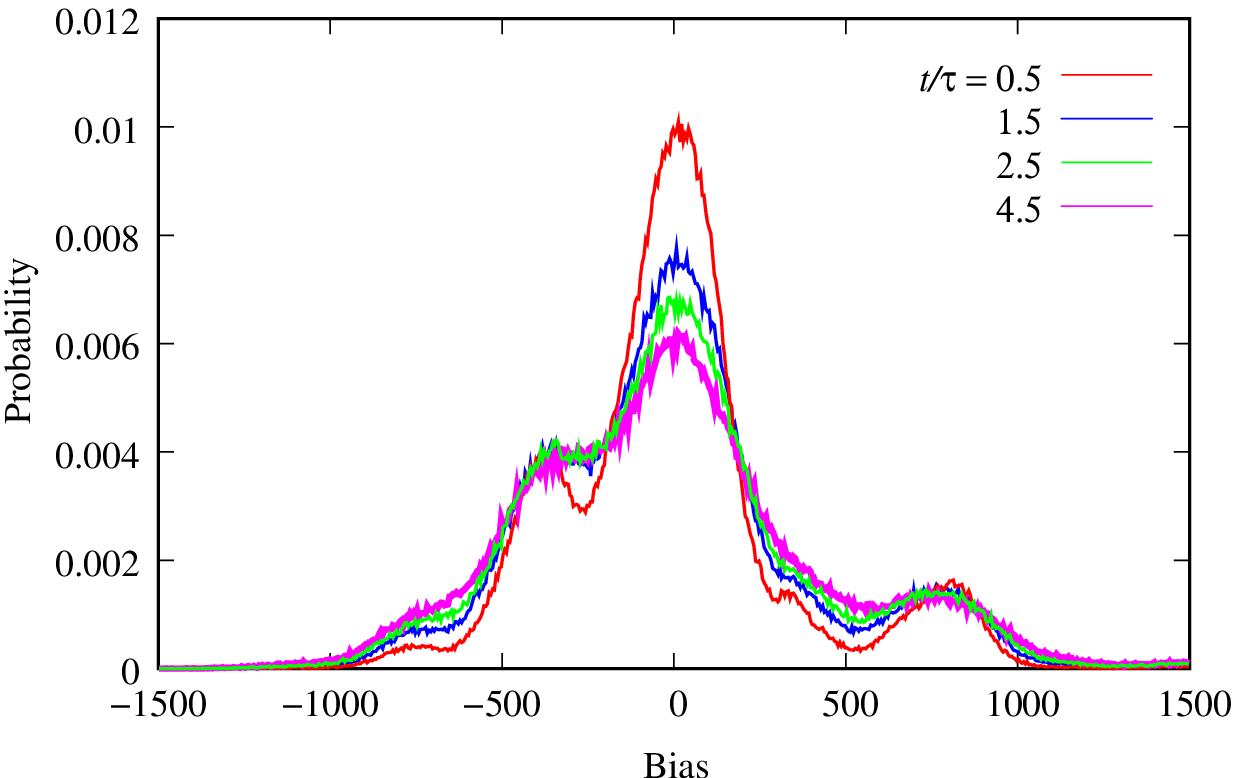}
\caption{\label{dist_short_9k.eps}
(Color online) Same as \fno{dist_short_82k.eps} for the $N=9171$ spin sample. The average is
over 60 runs.
} 
\vskip15pt
\includegraphics{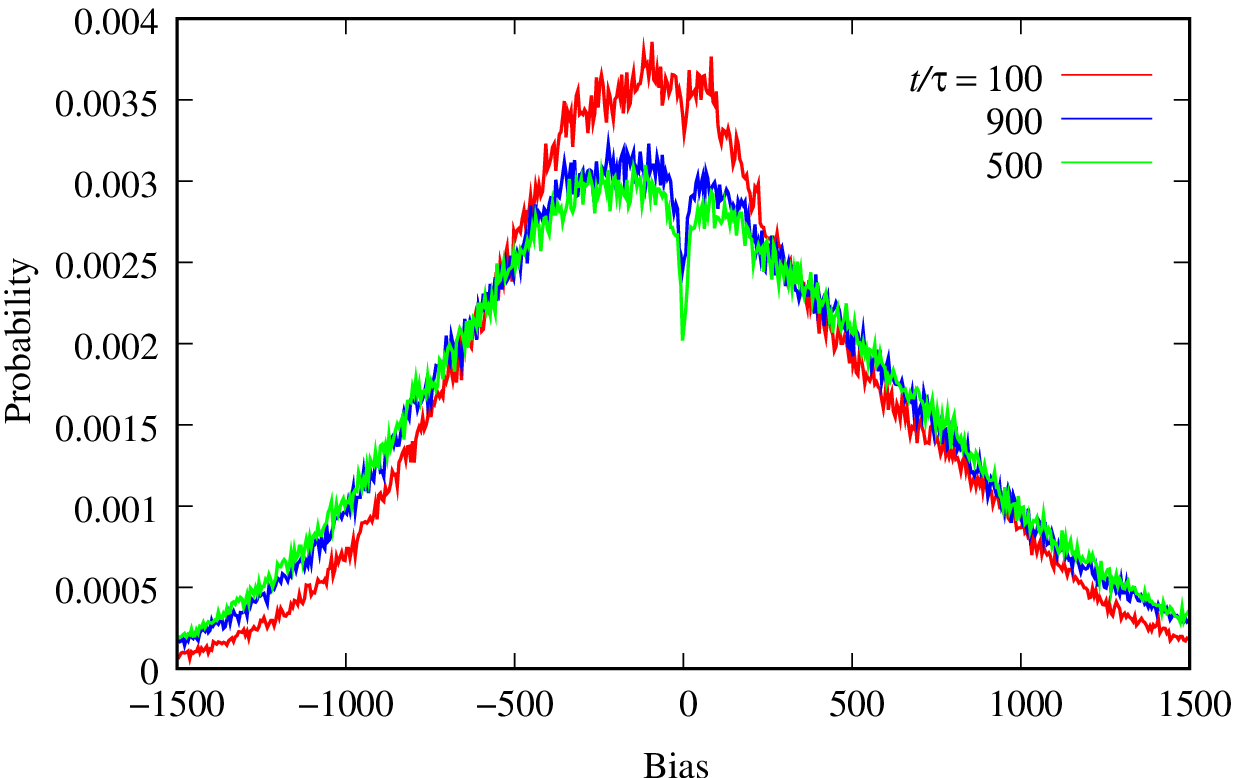}
\caption{\label{dist_long_9k.eps}
(Color online) Same as \fno{dist_long_82k.eps} for the $N = 9171$ sample. The average is
over 30 runs.
} 
\end{figure}
\begin{figure}
\includegraphics{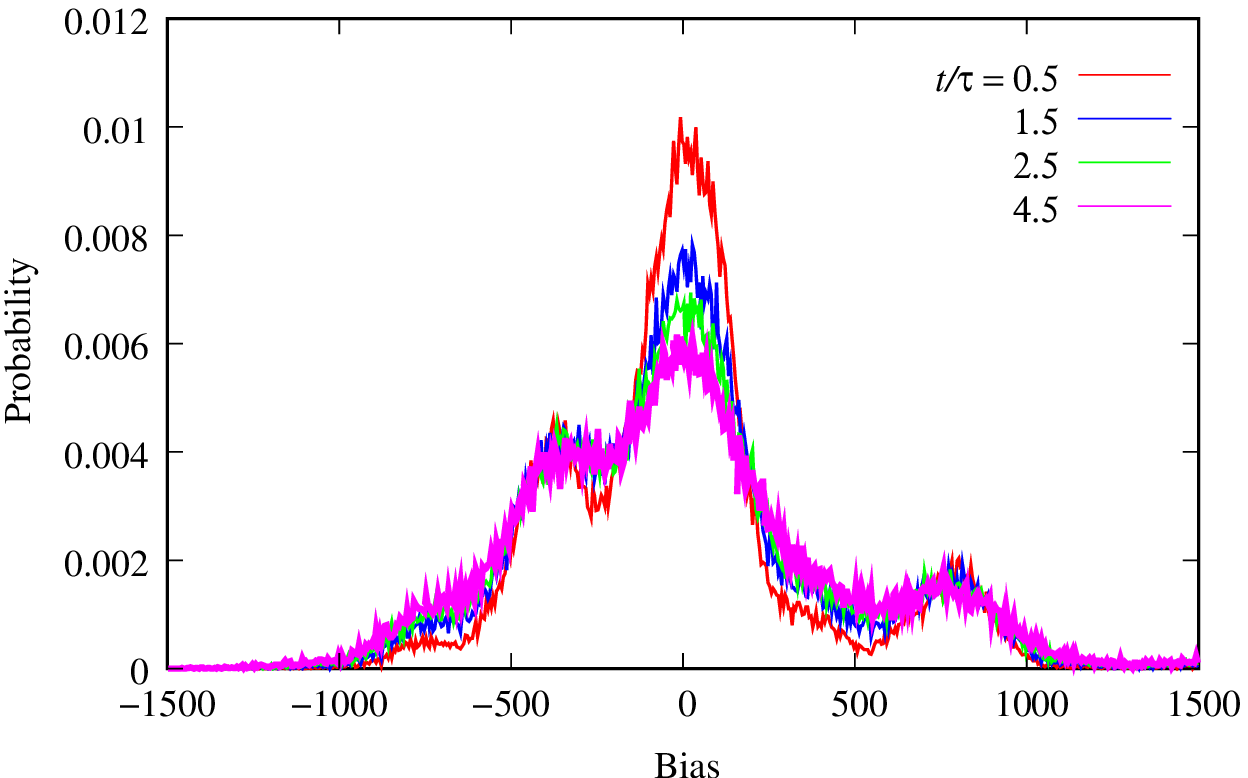}
\caption{\label{dist_one_shot_82k.eps}
(Color online) Single-run short-time bias distribution for the $N = 82519$ sample, with the
same parameters as before. There is no averaging.
} 
\end{figure}
%
%
\section{Rate Equations for Magnetization Relaxation}
\label{theory}
To understand our simulations, we have developed a theory based on rate equations.
The key realization lies in the very different role played by the reversible and the
nonreversible spins, and that we therefore need to understand the time-development of
each set separately. We denote by $N_r$, $N_{r\up}$, and $N_{r\dn}$ the total number
of reversible spins at any instant (i.e., those with a bias satisfying $|E| \le W$),
and the parts of this number whose spins are up or down. Corresponding lower case
symbols $n_r$, $n_{r\up}$, $n_{\up}$ etc. are used for the fractions $N_r/N$, $N_{r\up}/N$,
$N_{\up}/N$, etc. We also denote the number of nonreversible spins, $N - N_r$, by
$N_{\rb}$, and the sets of spins of various types by $\sr$, $\srup$, $\srb$ etc.
These sets obey obvious relations such as $\sr = \srup \cup \srdn$ and so on, which
need not be listed. It also pays to introduce the {\it reversible
magnetization\/},
\beq
m_r = n_{r\up} - n_{r\dn},
\eeq
the total magnetic moment $M = N m$, and its reversible part, $M_r = N m_r$.

\subsection{Processes that change the state of a spin}
\label{processes}
We now examine how different spins can develop in a small time interval $dt$.
A non-reversible spin (at site {\it i\/}, say) can

\begin{enumerate}
\item {\it Move into\/} the reversible bias range with a probability $\pin{i}$.
\item {\it Remain\/} in the non-reversible range with a probability $1 - \pin{i}$.

Naturally, since this spin cannot flip in the interval $dt$, these possibilities
depend on the behavior of other spins. We shall address the probability $\pin{i}$
below.
\end{enumerate}

A reversible spin (again taken to be at site {\it i\/}), on the other hand,
can do the following:

\begin{enumerate}
\item {\it Flip\/} and {\it move out\/} of the reversible range with probability
       $\pfl \pout{i}$.
\item {\it Flip\/} and {\it remain\/} in the reversible range with probability
       $\pfl (1- \pout{i})$.
\item {\it Not flip\/} and {\it become nonreversible\/} with probability
       $(1-\pfl) \pout{i}$.
\item {\it Not flip\/} and {\it stay reversible\/} with reversibility
       $(1 - \pfl) (1 - \pout{i})$.
\end{enumerate}

Once again, the probability $\pout{i}$ depends on the behavior of other spins, and will
be estimated below. We have also introduced the quantity
\beq
\pfl = \Gam_0 dt,
\eeq
in which the index {\it i\/} is omitted in $\pfl$, since this is the
flip probability for {\it all\/} reversible spins. Clearly, our model assumes that the
processes of flipping and of moving in or out of the reversibility range are independent,
which in turn means that different spins flip or do not flip completely independently of
each other, with a probability that depends only on the local bias. This assumption
will be valid provided the bias distribution $\rho(E)$ is reasonably spatially
homegeneous across the sample at all times. Such is the case for our spherical samples,
but will need to be reexamined for other shapes.

With the above hypothesis, the change in the numbers of various types of spins 
in a short time interval $dt$ are easily written down. For $dN_{r\up}$, we have
\beq
dN_{r\up}
  = -\sum_{i \in \srup}
       \Bigl[ \pfl\pout{i} + \pfl(1 - \pout{i}) + (1 - \pfl)\pout{i} \Bigr]
    + \sum_{i \in \srdn} \pfl(1 - \pout{i})
    + \sum_{i \in \srbup} \pin{i}.
\eeq
The first four terms on the right correspond to the four processes enumerated
above for reversible spins, while the fifth term corresponds to nonreversible
up spins becoming reversible. Simplifying, we get
\beq
dN_{r\up}
  = -\sum_{i \in \srup} \Bigl[ \pfl + (1 - \pfl)\pout{i} \Bigr]
    + \sum_{i \in \srdn} \pfl(1 - \pout{i})
    + \sum_{i \in \srbup} \pin{i}.                       \label{dNrup}
\eeq
Similarly,
\beq
dN_{r\dn}
  = -\sum_{i \in \srdn} \Bigl[ \pfl + (1 - \pfl)\pout{i} \Bigr]
    + \sum_{i \in \srup} \pfl(1 - \pout{i})
    + \sum_{i \in \srbdn} \pin{i}.                       \label{dNrdn}
\eeq
Adding the last two equations, we get a very simple equation for the change
in the total number of reversible spins,
\beq
dN_r = -\sum_{i \in \sr} \pout{i} + \sum_{i \in \srb} \pin{i},    \label{dNr}
\eeq
which does not depend on $\pfl$ at all, since we do not discriminate between
up and down spins in the set $\sr$, and the changes in its size are a function
of the behavior of neighboring spins of the members of this set.

By taking the difference of \etwo{dNrup}{dNrdn}, we get the change in the
unnormalized reversible magnetization:
\beq
dM_r = dN_{r\up} - dN_{r\dn}.
\eeq
We can simplify the expression that results upon substitution of the actual
forms of $dN_{r\up}$ and $dN_{r\dn}$ by anticipating that the probabilities
$\pout{i}$ and $\pin{i}$ will also be proportional to $dt$. Thus terms such as
$\pfl\pout{i}$ are $O(dt)^2$ and may be omitted. In this way, we get
\beq
dM_r = -2\pfl M_r
       - \Bigl(\sum_{i\in \srup} \pout{i} - \sum_{i\in\srdn} \pout{i} \Bigr)
       + \Bigl(\sum_{i\in \srbup} \pin{i} - \sum_{i\in\srbdn} \pin{i} \Bigr).
                   \label{dMr}
\eeq

Lastly, we find $dM$, the change in the total unnormalized magnetization. Since
this change can come about only by the flipping of reversible spins, and since
each flip changes $M$ by 2,
\bea
dM &=& -2 \sum_{i\in\srup} \pfl + 2 \sum_{i\in\srdn} \pfl \nnu \\
   &=& -2 \pfl M_r.                    \label{dM}
\eea

\subsection{The probabilites $\pinb$ and $\poutb$}
\label{pin_pout}

For the equations for $dN_r$, $dM_r$, and $dM$ to be useful, we need the
probabilities $\pinb$ and $\poutb$. Let us begin by considering a nonreversible
spin $i \in \srb$ that sees a bias $E_i > W$. For this spin to move into
the reversible range, {\it reversible\/} spins at other sites will need to
flip and alter the bias at site {\it i\/} to satisfy $|E'_i| \le W$, where
the prime indicates the bias after a time interval $dt$. Now,
\beq
E'_i = E_i + \sumpr_{\!\!j \in \sr} K_{ij} d\sig_j.  \label{bias_shift}
\eeq
Here, the site {\it i\/} is excluded from the sum, and $d\sig_j$ is the
change in the spin at site {\it j\/} in the time $dt$. The requirement that
$|E'_i| < W$ implies that only a particular set of reversible spins determined
by the geometry of the lattice and the form of the dipole kernel $K_{ij}$ can be
effective in making spin {\it i\/} reversible. We shall refer to such spins as
{\it triggering\/} spins. To estimate their number
we make the critical simplification that we may ignore
simultaneous spin flips since such processes will have a very low probability
proportional to $(dt)^2$, which may be neglected as $dt$ is infinitesimal.
Thus, in \eno{bias_shift}, we take $d\sig_j = 0$ for all but one distant
reversible spin. Taking this spin to be up, so that $d\sig_j = -2$, we get
\beq
E'_i = E_i - 2K_{ij}(\up),
\eeq
where the arrow in $K_{ij}(\up)$ indicates that the distant spin is up.
The condition $|E'_i| < W$ then implies that
\beq
{E_i - W \by 2} \le K_{ij}(\up) \le {E_i + W \by 2}.  \label{K_range_1}
\eeq
Similarly, if the distant spin is down, we require
\beq
{- E_i - W \by 2} \le K_{ij}(\dn) \le {- E_i + W \by 2}.  \label{K_range_2}
\eeq

We now find the number of sites for which the couplings $K_{ij}$ lie in the
range (\ref{K_range_1}) or (\ref{K_range_2}). If we define
\beq
K_1(E_i) = \hf (E_i - W), \quad K_2(E_i) = \hf (E_i + W),   \label{K1K2}
\eeq
then these two ranges correspond to intervals $[K_1, K_2]$, and
$[-K_2, -K_1]$ in which $K_{ij}$ must lie. Let us denote the numbers of
sites in each interval by $N_{[K_1, K_2]}$ and $N_{[-K_2, -K_1]}$. We have
\beq
N_{[K_1, K_2]} = \int_{K_1}^{K_2} g(K)\, dK,
\eeq
and similarly for $N_{[-K_2, -K_1]}$, where $g(K)$ is the density of
dipole couplings found in Ref.~\cite{avag09}. That is, $g(K) dK$ is the
number of sites for which the coupling to a central site lies between $K$
and $K + dK$. We have
\beq
g(K) = \al {E_{dm} \by K^2}, \quad \al = {16 \pi \by 9\sqrt{3}}.
\eeq
It then follows that
\bea
N_{[K_1, K_2]} &=& N_{[-K_2, -K_1]} \nnu \\
               &=& \al E_{dm}\biggl({1\by K_1} - {1\by K_2} \biggr)
                        \label{Nk1k2} \\
               &=& 4 \al {E_{dm} W \over E_i^2 - W^2}. \label{Nk1k2_final}
\eea
Since the number of distant sites at which a triggering spin could be located
is independent of whether that spin is up or down, we can calculate the probability
that spin $i$ will become reversible, that is to say $\pin{i}$, as the product of
three factors: (i) the number (\ref{Nk1k2_final}), (ii) the fraction of these sites
at which the spin is itself reversible, $n_r$, and (iii) the probability that any one
of these spins will flip, $\pfl$. Thus,
\beq
\pin{i} = 4 \al n_r {E_{dm} W \over E_i^2 - W^2} \Gam_0 dt. \label{pin_i}
\eeq
The above calculation assumes once again that the local reversible fraction
$n_r$ in the vicinity of spin {\it i\/} is spatially homogeneous, and thus
independent of the location of site {\it i\/}.

We next turn to the calculation of $\poutb$, which proceeds in close
parallel to that of $\pinb$. Consider a reversible spin at
site {\it i\/}, i.e., the bias $E_i$ obeys $|E_i| \le W$. We refer to this
site as the central reversible spin. This spin will become nonreversible
if a distant reversible spin flips in such a way as to push the bias
at site {\it i\/} outside the interval $[-W, W]$. Suppose the distant spin
flips from up to down. Since we have $|E_i| \le W$ and want $|E'_i| > W$,
the coupling $K_{ij}$ must be such that
\beq
K_{ij} \not\in [K_1, K_2],
\eeq
where $K_1$ and $K_2$ are as defined in \eno{K1K2}. Noting that now
$K_1 < 0$ and $K_2 > 0$, the number of sites that meet this requirement is
given by
\beq
\int_{-\infty}^{K_1} g(K)\,dK + \int_{K_2}^{\infty} g(K)\, dK
  = 4 \al {E_{dm} W \by W^2 - E_i^2}. \label{N_dist_up}
\eeq
Similarly, if the distant spin flips from down to up, the condition on
$K_{ij}$ is
\beq
K_{ij} \not\in [-K_2, -K_1],
\eeq
which is met by a number of sites equal to
\beq
\int_{-\infty}^{-K_2} g(K)\,dK + \int_{-K_1}^{\infty} g(K)\, dK
  = 4 \al {E_{dm} W \by W^2 - E_i^2}, \label{N_dist_dn}
\eeq
which is the same as \eno{N_dist_up}. Thus, once again, the number of
sites on which a triggering spin can be located is independent of whether
that spin is up or down, and we may calculate $\pout{i}$ as the
product of (i) the number of sites (\ref{N_dist_up}), (ii) the
fraction $n_r$ that the spin on one of these sites is reversible, and
(iii) the probabibility $\Gam_0 dt$ that this spin will indeed flip. Thus,
\beq
\pout{i} = 4 \al n_r {E_{dm} W \over W^2 - E_i^2} \Gam_0 dt. \label{pout_i}
\eeq

The expressions (\ref{pin_i}) and (\ref{pout_i}) suffer from unpleasant
singularities when $E_i = \pm W$. These singularities are unphysical, and
are a consequence of using the modified spin-flip probabillity
(\ref{p_flip2}) with the hard cutoffs at $\pm W$. Better estimates are
obtained by noting that for $\pin{i}$, $|E_i|$ is likely to be much bigger
than $W$, while for $\pout{i}$ the converse is true. We therefore neglect
the term $W^2$ in the denominator of \eno{pin_i} and $E_i^2$ in the
denominator of \eno{pout_i}, leading to the expressions
\bea
\pinb(E_i) &=& 4 \al n_r {E_{dm} W \over E_i^2} \Gam_0 dt, \label{pin} \\
\poutb(E_i) &=& 4 \al n_r {E_{dm} \over W} \Gam_0 dt. \label{pout}
\eea
We note here the intuitively reasonable fact that $\poutb$ is much greater
than $\pinb$. The set $\sr$ is much smaller than $\srb$, so an initially
reversible spin will be knocked out of reversibility by almost all flips of
neighboring spins. By contrast, to move an initially nonreversible spin into
reversibility, one must cancel the preexisting bias at the nonreversible site
nearly exactly, which can only be done by flipping distant spins at a very
specific set of sites. For this same reason, $\pout{i}$ essentially
does not depend on $E_i$, while $\pin{i}$ does.

Note also that $\pinb$ and $\poutb$ are both proportional to $dt$ as anticipated
earlier.
\subsection{The rate equations}
\label{rate_eqns}
We now substitute \etwo{pin_i}{pout_i} into Eqs.~(\ref{dNr}), (\ref{dMr}), and
(\ref{dM}) for $dN_r$, $dM_r$, and $dM$, and divide by the total number of spins
$N$ at the same time in order to get equations for intensive quantities. Let us
begin by considering the two sums in \eno{dNr} one by one. Since $\pout{i}$ is
independent of $E_i$ as noted above, we have
\beq
{1\by N} \sum_{i \in \sr} \pout{i}
   = \poutb {N_r \by N} 
   =  4 \al n^2_r {E_{dm} \over W} \Gam_0 dt.
\eeq
For the second sum, we need to sum over the set $\srb$. We do this by including
all sites where the bias exceeds $W$ in magnitude. This leads to the approximation 
\beq
{1\by N} \sum_{i \in \srb} \pin{i} = 4 \al n_r {E_{dm} \over W} {\cal F} \Gam_0 dt, 
\eeq
where ${\cal F}$ is a dimensionless functional of the bias distribution $\rho(E)$,
given by
\beq
{\cal F}[\rho(E)] = W^2 \int_{|E| > W} {\rho(E) \by E^2} dE. \label{defF}
\eeq
Hence,
\beq
{dn_r \by dt} = - 4 \al\Gam_0 {E_{dm} \by W} n_r (n_r - {\cal F}).   \label{dnrdt}
\eeq

Next, we examine \eno{dMr} for $dM_r/dt$. For the term with the sums over the sets
$\srup$ and $\srdn$, we have,
\beq
{1\by N} \Bigl(\sum_{i\in \srup} \pout{i} - \sum_{i\in\srdn} \pout{i} \Bigr)
   = 4 \al n_r m_r{E_{dm} \over W} \Gam_0 dt.
\eeq
For the remaining two sums, we estimate the sizes of the sets $\srbup$ and $\srbdn$
as $N_{\up}$ and $N_{\dn}$ times the size of $\srb$ on the theory that when $n_r \ll 1$,
most of the spins are nonreversible and the bias at any site is uncorrelated with
whether the spin at that site is up or down, and that when $n_r \simeq 1$, $m \simeq m_r$.
It follows that
\beq
{1\by N} \Bigl(\sum_{i\in \srbup} \pin{i} - \sum_{i\in\srbdn} \pin{i} \Bigr)
   \simeq 4 \al n_r m {E_{dm} \over W} {\cal F} \Gam_0 dt.
\eeq
Hence,
\beq
{dm_r \by dt} = - 2\Gam_0 m_r - 4 \al\Gam_0 {E_{dm} \by W} n_r (m_r - m {\cal F}).
             \label{dmrdt}
\eeq

Lastly, we obtain the equation for $dm/dt$, which is the simplest of all:
\beq
{dm\by dt} = -2\Gam_0 m_r.   \label{dmdt}
\eeq

Equations (\ref{dnrdt}), (\ref{dmrdt}), and (\ref{dmdt}) are the desired rate equations.
They are manifestly nonlinear, but more importantly and contrary to our initial hope,
they are not a closed system because of the presence of the functional ${\cal F}$ of the
full bias distribution $\rho(E)$. At present this puts a big limitation on their use. For
the relaxation problem we have been able to circumvent this limitation by constructing an
interpolation form for $\rho(E)$ which we believe is reasonably accurate and
self-consistent over a wide range of times, well past that over which the square-root time
development is seen. We describe our approximation for $\rho(E)$ in the next section.
\section{The bias distribution}
\label{bias_dist}
\subsection{The three-Gaussian approximation}
\label{tga}
As seen from the Monte Carlo simulations, the bias distribution at short times is
dominated by three peaks at $E=0$, $E = -4E_{dm}$, and $E= 8E_{dm}$. The locations of the
two side peaks are a strong indicator of their origin. Consider a site with its six nearest
neighbors. Four of these neighbours are in the \itxy plane, and two are along the \itz
axis. If any of the neighboring spins in the \itxy plane flips from up to down, the bias
at the central site will change by an amount $-4E_{dm}$, while if any of the \itz axis
neighbours flips, the field at the central site will change by $8 E_{dm}$. This explains
the peak locations. Further, since there are twice as many near neighbours of any site
in the \itxy plane as there are along the \itz axis, we should expext the peak at
$-4E_{dm}$ to be about twice as high as the peak at $8 E_{dm}$ as long as $N_r \ll N$.
This is also seen in the data. The smaller peak at $-8E_{dm}$ and shoulder at
$4E_{dm}$ can also be associated with spin flips at pairs of near neighbour sites.

Motivated by this idea, we try and represent $\rho(E)$ as a sum of three Gaussians
centered at 0, $-4 E_{dm}$, and $8 E_{dm}$. Suppose that at a given time, $N_{\dn}$
spins have flipped where $N_{\dn} \ll N$, allowing us to ignore the possibility that
two flipped spins are near neighbours of each other or even of a common third spin.
Then there are $4N_{\dn}$ spins that have a flipped neignbour in the \itxy plane, and
$2N_{\dn}$ spins that have a flipped neigbour along the \itz axis, leaving
$N - 6N_{\dn}$ spins which have no flipped neighbours at all. Thus the weights of
the 0, $-4E_{dm}$ and $8E_{dm}$ peaks are proportional to $(1- 6 n_{\dn})$,
$4 n_{\dn}$, and $2 n_{\dn}$ respectively. We can further argue that the widths of
all three peaks are equal and proportional to $n_{\dn}^{1/2}$, since the fields at
sites far away from all flipped spins should continue to vanish on average, but
should have a variance that grows linearly with the number of
flipped spins. For a site next to a flipped spin, this variance is simply realized
around the shift produced by the flipped neighbour. Thus for $n_{\dn} \ll 1$, the
three-Gaussian approximation (TGA) to $\rho(E)$ takes the form
\beq
\rho(E) \simeq (1 - 6 n_{\dn}) g_0(E) + 2 n_{\dn} g_{+}(E) + 4 n_{\dn} g_{-}(E),
      \label{rho_small_n}
\eeq
where (with $\al = 0$, $+$, or $-$, and $E_0 =0$, $E_{+} = 8 E_{dm}$, and
$E_{-} = - 4E_{dm}$)
\beq
g_{\al}(E) 
   = (2\pi n_{\dn} \sigt^2)^{-1/2} e^{-(E - E_{\al})^2/2 n_{\dn} \sigt^2},
     \quad (n_{\dn} \ll 1). \label{g_small_n}
\eeq
The quantity $\sigt$ is $E_{dm}$ times an unknown constant of order unity.

The arguments underlying \eno{rho_small_n} start to become questionable for $n_{\dn}$
as small as $0.1$, since sites with two near neighbour flipped spins start to become
significant. To enable us to consider larger values of $n_{\dn}$, we generalize the
TGA to the form
\beq
\rho(E) \simeq a_0 g_0(E) + a_{+} g_{+}(E) + a_{-} g_{-}(E),
      \label{rho_gen}
\eeq
where
\beq
g_{\al}(E) = (2\pi \sig^2)^{-1/2} e^{-(E - E_{\al})^2/2 \sig^2}. \label{g_gen}
\eeq
That is, the peaks of the three Gaussians are still taken to be at $0$, $-4E_{dn}$
and $8 E_{dn}$, the widths are taken to have a common value $\sig$ not necessarily
proportional to $n_{\dn}^{1/2}$, and the weights $a_0$, $a_{+}$, and $a_{-}$ are
allowed to become arbitrary. We will determine these weights and the
width by the procedure described in the next subsection. The form (\ref{rho_small_n})
at small $n_{\dn}$ will serve as a check on the procedure.

It is apparent that the TGA is qualitatively incapable of accounting for the very
narrow hole that is burned in the distribution at long times, but here a different
approximation scheme can be developed as the origin of the hole is physically
obvious.
\subsection{Moments of the bias distribution for uncorelated spins}
\label{moments}
Our discussion above implies that for very small $n_{\dn}$, the flipped spins are
randomly distributed in the lattice without any spatial correlations. We therefore
extend this idea to larger $n_{\dn}$ and consider a model in which the spin on each
site is up or down independently of other spins, with probabilities $(1 \pm m)/2$,
where $m$ is the magnetization. We then calculate the first three moments of this
model, and match those to the moments of the TGA, \eno{rho_gen}. These three
moments, plus the normalization (or zeroth moment) give us the four conditions
needed to determine the four quantities $a_0$, $a_{-}$, $a_{+}$, and $\sig$.

The bias at any site \iti is given by
\beq
E_i = \sum_{j\ne i} K_{ij} \sig_j.
\eeq
Consider first the uniform spin configuration with $m = 1$, i.e., $\sig_i = 1$ for
all {\it i\/}.  We know that in this case the bias vanishes at all sites except those in
a narrow layer near the surface of our spherical sample. Hence we may take
\beq
\sum_{j\ne i} K_{ij} = 0     \label{sumK}
\eeq
for essentially all sites. This result will be employed repeatedly in the
calculations of the moments for configurations in which $m \ne 1$. Thus, for the first
moment, we have
\bea
\avg{E_i} &=& \sum_{j\ne i} K_{ij} \avg{\sig_j} \nnu \\
          &=& \sum_{j\ne i} K_{ij} m \nnu \\
          &=& 0.                               \label{avgE}
\eea

Similarly, for the second moment, we get
\beq
\avg{E_i^2} = \sumpr_{\!\!j, k} K_{ij} K_{ik} \avg{\sig_j \sig_k}.
\eeq
The prime on the sum signifies that $j \ne i$ and $k \ne i$.
Now $\avg{\sig_j \sig_k}$ equals 1 if $j = k$, and $m^2$ if $j \ne k$. Hence,
\bea
\avg{E_i^2} &=& \sumpr_{\!\!j, k} K_{ij} K_{ik}
                    [\dta_{jk} + (1 - \dta_{jk}) m^2]  \nnu \\
            &=& \sum_{j\ne i} K^2_{ij} (1 - m^2) 
                   + m^2 \sumpr_{\!j} K_{ij} \sumpr_{\!k} K_{ik} \nnu \\
            &=& \kap_2 E_{dm}^2 (1 - m^2),   \label{avgE2}
\eea
where we have used \eno{sumK}, and defined
\beq
\kap_2 = {1\by E_{dm}^2} \sum_{j \ne i} K^2_{ij}.
\eeq
Numerical evaluation of the sum gives
\beq
\kap_2 = 53.427   \label{kap2}
\eeq

For the third moment, we have 
\beq
\avg{E_i^3} = \sumpr_{\!\!j, k, l} K_{ij} K_{ik} K_{il} \avg{\sig_j \sig_k \sig_l}.
\eeq
Again, the prime signifies that $j \ne i$, $k \ne i$, and $l \ne i$. The only issue
requiring care in performing the sum is the enumeration of the various cases of equality
or inequality of the indices $j$, $k$, and $l$. The first case is where all three
indices are distinct. Then $\avg{\sig_j \sig_k \sig_l} = m^3$, and the contribution
of this case to $\avg{E^3_i}$ can be evaluated as
\bea
{\avg{E^3_i}}_1
  &=& m^3 \sumpr_{\!\!j, k, l} K_{ij}K_{ik}K_{il}
                   (1 - \dta_{jk}) (1 - \dta_{kl}) (1 - \dta_{lj}) \nnu \\
  &=& m^3 \sumpr_{\!\!j, k, l} K_{ij}K_{ik}K_{il}
                   (1 - 3\dta_{jk} + 3 \dta_{jk} \dta_{jl}
                                  - \dta_{jk}\dta_{kl}\dta_{lj})   \nnu \\
  &=& m^3 \biggl[ \Bigl(\sumprj K_{ij} \Bigr)^3
                          - 3\sumprj K_{ij}^2 \sumpr_{\!\!l} K_{il}
                              + 3\sumprj K_{ij}^3
                                  - \sumprj K_{ij}^3     \biggr]  \nnu \\
  &=& 2m^3 \sum_{j\ne i} K^3_{ij}.
                                   \label{Ecube1}
\eea
In line 2 above we have used the symmetry of the summand, and in line 4 we
have used \eno{sumK}.

The second case is where two of the indices $j$, $k$, and $l$ are the same, but
distinct from the third. Now $\avg{\sig_j \sig_k \sig_l} = m$. This case has three
identically contributing  subcases, and for its net contribution to $\avg{E^3_i}$
we have
\bea
{\avg{E^3_i}}_2
  &=& 3m \sumpr_{\!j, k, l} K_{ij}K_{ik}K_{il}\,
                         \dta_{jk} (1 - \dta_{jl}) \nnu \\
  &=& 3m \biggl[ \sumprj K^2_{ij} \sumpr_{\!\!k} K_{ik}
                    -\sumprj K^3_{ij} \biggr]      \nnu \\
  &=& -3m \sum_{j\ne i} K^3_{ij},
                                   \label{Ecube2}
\eea
where we have again used \eno{sumK} in the last line.

The third and last case is that where $j = k = l$. Now
$\avg{\sig_j \sig_k \sig_l} = m$, and the contribution to $\avg{E^3_i}$ is,
therefore,
\beq
{\avg{E^3_i}}_3 = m \sum_{j\ne i} K^3_{ij}.  \label{Ecube3}
\eeq

Adding together Eqs.~(\ref{Ecube1}), (\ref{Ecube2}), and (\ref{Ecube3}), we get
\beq
{\avg{E^3_i}} = -2m (1 - m^2) \sum_{j\ne i} K^3_{ij}.
\eeq
We write this as
\beq
{\avg{E^3_i}} = \kap_3 E_{dm}^3 m (1 - m^2),   \label{avgE3}
\eeq
where
\beq
\kap_3 = -{2\by E_{dm}^3} \sum_{j\ne i} K^3_{ij} = 190.47, \label{kap3}
\eeq
and the last result is found numerically.

It should be noted that in this model, the moments of $E$ are simply geometrical
constants determined by the type of lattice times the appropriate power of the
energy scale $E_{dm}$.
\subsection{Moment matching}
\label{match_mom}
We now match the moments from the previous subsection with those of the
three-Gaussian approximation (\ref{rho_gen}). The latter yields
\bea
\avg{E} &=& 8E_{dm} a_{+} - 4E_{dm} a_{-}, \label{tgaE1} \\
\avg{E^2} &=& \sig^2(a_0 + a_{+} + a_{-}) 
                + 64 E_{dm}^2 a_{+} + 16 E_{dm}^2 a_{-}, \label{tgaE2} \\
\avg{E^3} &=& 12 E_{dm} \sig^2 (2 a_{+} - a_{-} )
                + 512 E_{dm}^3 a_{+} - 64 E_{dm}^3 a_{-}. \label{tgaE3}
\eea
Equating these moments to those from the uncorrelated spin distribution yields
\bea
4 E_{dm} (2 a_{+} - a_{-}) &=& 0, \\
\sig^2 (a_0 + a_{+} + a_{-}) 
    + 16 E_{dm}^2 (4 a_+ + a_{-}) &=& \kap_2 E_{dm}^2 (1-m^2), \\
12 E_{dm} \sig^2 (2 a_+ - a_{-})
    + 64 E_{dm}^3 (8 a_+ - a_{-}) &=& \kap_3 E_{dm}^3 m(1-m^2).
\eea
Solving these equations along with the normalization condition,
\beq
a_0 + a_{+} + a_{-} = 1, \label{tgaE0}
\eeq
we obtain
\bea
a_0 &=& 1 - {\kap_3 \by 128} m (1-m^2), \label{a0}\\
a_{+} &=& {\kap_3 \by 384} m (1-m^2), \\
a_{-} &=& {\kap_3 \by 192} m (1-m^2), \\
\sig^2 &=& {1\by 4} (4\kap_2 - \kap_3 m) (1-m^2) E_{dm}^2. \label{sig_sq}
\eea

At this point let us ask whether the solution (\ref{a0})--(\ref{sig_sq}) 
approaches \etwo{rho_small_n}{g_small_n} when $n_{\dn} \ll 1$. In that limit,
since $m = 1 - 2 n_{\dn}$, $m(1-m^2) \apx (1-m^2) = 4n_{\dn}$. Feeding in
the value $\kap_3 = 190.2$, we get $a_{+} = 1.98 n_{\dn}$, and
$a_{-} = 3.96 n_{\dn}$, instead of $2n_{\dn}$ and $4n_{\dn}$. The differences
are rather small, however, and can be eliminated entirely if we make
the replacement
\beq
\kap_3 \to \kap'_3 = 192.
\eeq
This leads to the final forms we shall use in our three-Gaussian
approximation, \etwo{rho_gen}{g_gen}:
\bea
a_0   &=& 1 - {\tst{3\by2}} m (1-m^2), \label{a0fin}\\
a_{+} &=& \tshf m (1-m^2), \\
a_{-} &=& m (1-m^2), \\
\sig^2 &=& (\kap_2 - 48 m) (1-m^2) E_{dm}^2. \label{sig_sq_fin}
\eea
\subsection{Comparison with simulations}
\label{compar_tga}
When we now compare the TGA with the simulations, we discover that the agreement is off
by $\sim 10\%$ if we use the value $\kap_2 = 53.4$. This value was calculated for an
infinite lattice, and for a finite sized sample the variance of $E_i^2$ should be smaller.
Using the value 50 appropriate to the 82519 spin sample, we find that the agreement is
considerably improved. In \fno{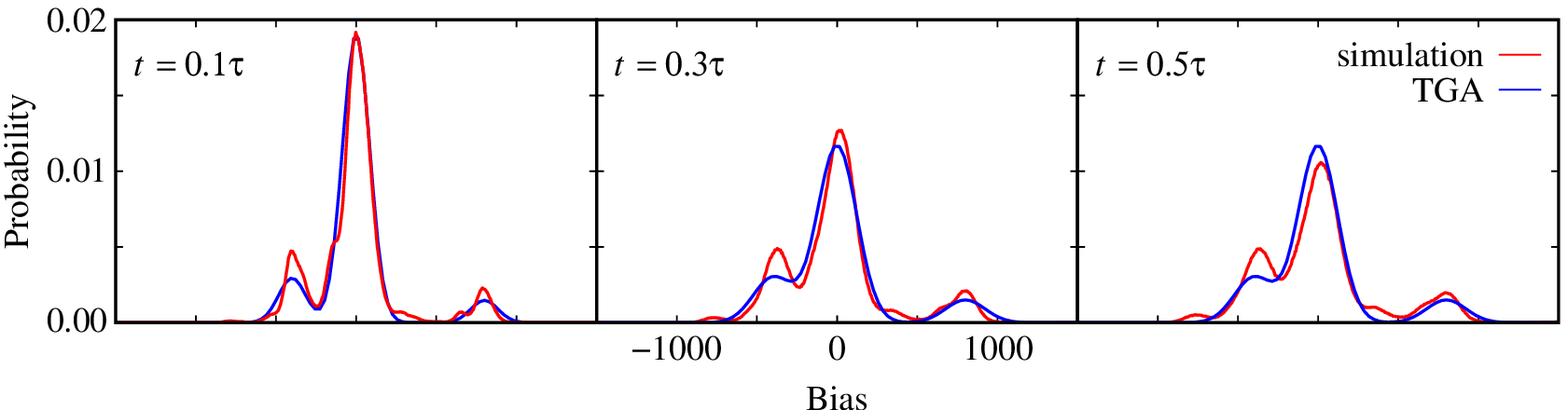} we show the TGA with the choice $\kap_2 = 50$
along with the results of the simulations for
the $82519$ spin sample for $t/\tau = 0.1$, $0.3$, and $0.5$, where
$\tau = E_{dm}/\Dta^2$.  At these three times, $m = 0.93$, $0.89$, and $0.86$.
The agreement becomes poorer for larger $t$, and it is about as good as could be expected
given how simple-minded the approximation is.
\begin{figure}
\includegraphics{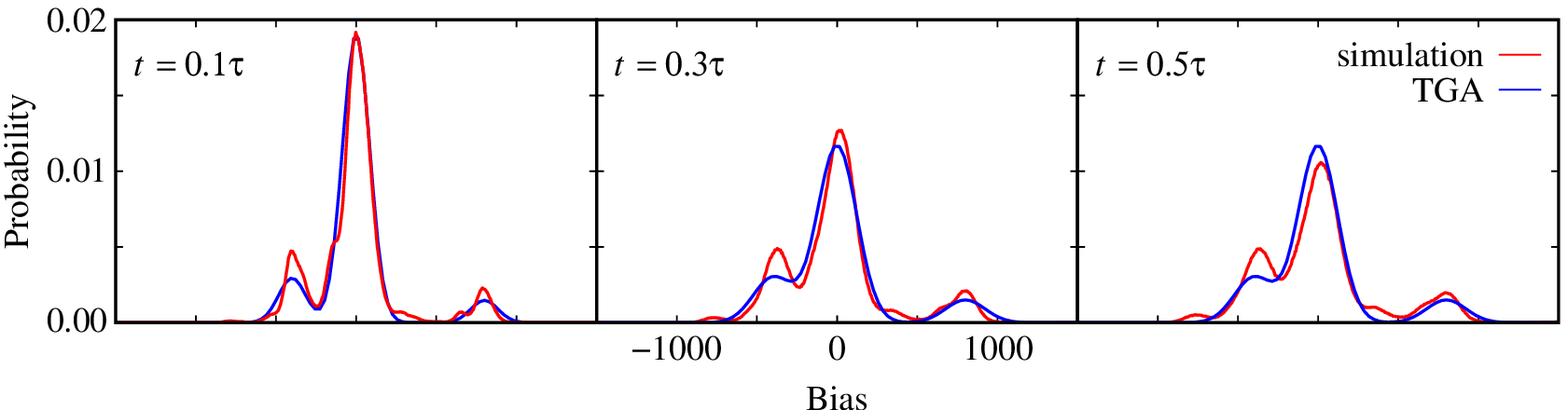}
\caption{\label{tga_fit_all.eps}
(Color online) Comparison between the three-Gaussian approximation (TGA) to the bias
distribution and the simulation results for short times. The sample has
$N = 82519$ spins, and all other parameters are as in previous figures.
} 
\end{figure}
\section{Short-time decay of magnetization: the $\sqrt{t}$ law}
\label{root_t}
In \fno{m_vs_t_short.eps} we show $m(t)$ for short times from our simulations, and from
solving the rate equations with the value $\kap_2 = 53.4$. As can be seen the general trend
is the same, although the detailed agreement is only good to about 3\%. Once again, the agreement
is improved if we set $\kap_2 = 50$, as shown in \fno{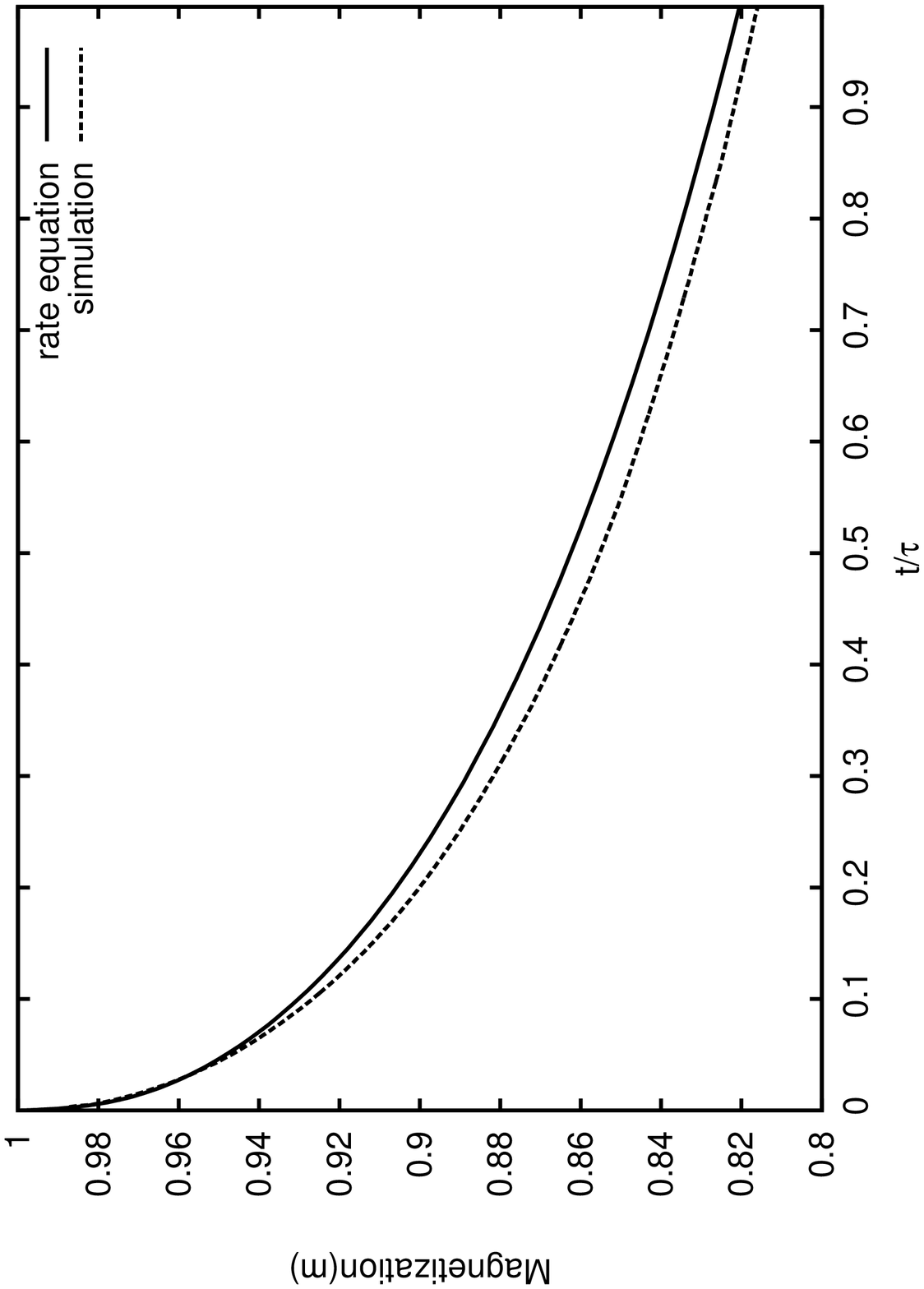}. The same data are shown on
a log-log plot in \fno{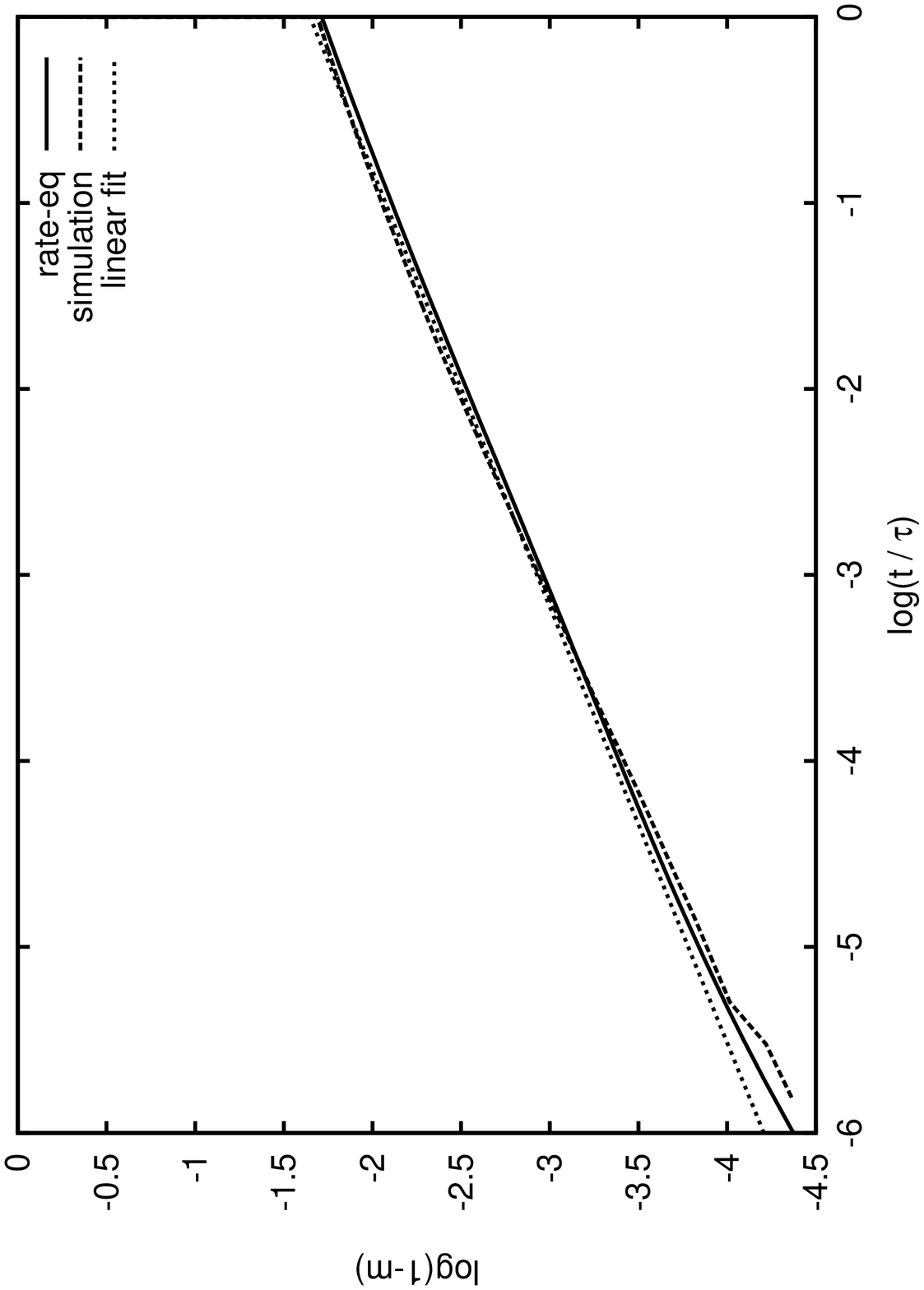}. As can be seen, both the simulations and the rate equation
show a power law behavior, with the same exponent. The best fit gives an exponent of 0.46, which
is very close to 0.5 as it would be for $\sqrt{t}$ behavior. We now show that this behavior can
be understood analytically on the basis of our rate equations, and that this exponent does not
depend on the choice of $\kap_2$.

The first key point is that starting from a delta-function at $t=0$, the bias
distribution becomes broader than the reversibilty region at some ultra-short
time when the fraction of flipped spins is still very small. From \eno{sig_sq}, we find
that for $n_{\dn} \ll 1$,
\beq
\sig^2 \apx A^2 E_{dm}^2 n_{\dn},
\eeq
where $A^2 = 4\kap_2 - \kap_3$. Thus $\sig \ltwid W$ only as long as
$n_{\dn} \ltwid (W/A E_{dm})^2$, which is of order $10^{-3}$. For such ultrasmall values of
$n_{\dn}$, $n_r = 1 - 6n_{\dn}$, and ${\cal F} \simeq 0$, so the rate equation for
$n_r$ simplifies to
\beq
{d n_{\dn} \by dt} = {2\by 3} \al \Gam_0 {E_{dm}\by W}.
\eeq
This has the solution $n_{\dn} = (2\al\Gam_0 E_{dm}/3W)t$, and so the condition that
$\sig \ltwid W$ holds only for $t \ltwid \tus$, where
\beq
\tus \sim {1\by A^2} \left({W\by E_{dm}} \right)^4 \tau
\eeq
is an ultra-short time scale of order $10^{-5} \tau$.

It follows that there is a large range of times, $\tus \ltwid t \tau$, for which
$\sig \gg W$ even though $n_{\dn} \ll 1$, i.e., very few spins are flipped. Thus almost
all the weight in the bias distribution is still in the central Gaussian, i.e.,
$a_0 \apx =1$, and the dimensionless functional that determines the repopulation of the
reversibility region can be approximated as
\beq
{\cal F} = 2 {W^2 \by \sqrt{2\pi \sig^2}} \int_W^{\infty} {e^{-E^2/2\sig^2} \by E^2} dE.
\eeq
Now, by integrating by parts, we get
\bea
\int_W^{\infty} {e^{-E^2/2\sig^2} \by E^2} dE
  &=& {1\by W} e^{-W^2/2\sig^2} -  {1\by \sig^2} \int_W^{\infty} e^{-E^2/2\sig^2} dE \nnu \\
  &=& {1\by W} e^{-W^2/2\sig^2}
         - {1\by \sig^2}\left[\sqrt{{\pi\by 2}}\sig - \int_0^W e^{-E^2/2\sig^2} dE \right].
\eea
The last expression can be expanded in powers of $W$, and we get
\beq
{\cal F} \simeq \sqrt{{2\by\pi}} {W\by \sig}
                 \left(1 - \sqrt{{\pi\by 2}}{W\by\sig} + \cdots \right).
                                  \label{Fvssig}
\eeq

The second key point is that even though $n_{\dn} \ll 1$, almost all the spins have been
knocked out of the reversibility region, i.e., $n_r \ll 1$. To see this we again approximate
the bias distribution by neglecting the weight outside the central Gaussian, and setting
$a_0 = 1$, so
\beq
n_r \simeq {1 \by \sqrt{2\pi\sig^2}} \int_{-W}^W e^{-E^2/2\sig^2} dE.
\eeq
Expanding the integrand in powers of $E$ and integrating, we get
\beq
n_r \simeq \sqrt{{2\by\pi}} {W\by \sig}
                 \left(1 - {W^2\by 6\sig^2} + \cdots \right).
                                    \label{nrvssig}
\eeq
Thus, to first order in $W/\sig$, $n_r = {\cal F}$, and the difference is of higher order:
\beq
n_r - {\cal F} = {W^2 \by \sig^2}.
\eeq
We can express this in terms of $n_r$ itself by using \eno{nrvssig}. We have
\beq
{W \by \sig} \simeq \sqrt{\pi\by2} n_r,
\eeq
so
\beq
n_r - {\cal F} = {\pi \by 2} n_r^2.
\eeq
The rate equation for $n_r$ then reads
\beq
{dn_r \by dt} = -{\pi \by 2} \zeta n_r^3, \label{nreqn}
\eeq
where we have defined
\beq
\zeta = 4\al\Gam_0 {E_{dm}\by W} = \al {\Dta_2^2 E_{dm} \by W^2}.
\eeq

The integration of \eno{nreqn} is elementary. Since this equation only holds for
$t \gtwid \tus$, we can write the integral in the form
\beq
{1\by n_r^2} = \pi\zeta (t + t^*),
\eeq
where $t^*$ is a time of order $\tus$. We thus have an explicit solution for the time
dependence of the reversible fraction:
\beq
n_r(t) = {1\by \sqrt{\pi\zeta}}{1 \by (t+t^*)^{1/2}}.
\eeq
The other rate equations can now be solved as follows. We have by definition,
\beq
n_r = n_{r\up} + n_{r\dn}, \quad m_r = n_{r\up} - n_{r\dn}.
\eeq
Since $n_{r\dn} < n_{\dn} \ll 1$, the answer for $m_r$ is immediate:
\beq
m_r \apx n_r = {1\by \sqrt{\pi\zeta}}{1 \by (t+t^*)^{1/2}}.
\eeq
The rate equation for $m$ now reads
\beq
{dm\by dt} = -{2\Gam_0 \by \sqrt{\pi\zeta}} {1\by (t+t^*)^{1/2}}.
\eeq
The integration is again elementary. Assuming that $\tus \ltwid t \ltwid \tau$, we can write
the result as
\beq
m(t) \simeq 1 - \sqrt{\Gam_{1/2} t}, \label{mvst}
\eeq
where
\beq
\Gam_{1/2} = 16 {\Gam_0^2 \by \pi \zeta} = {1\by \pi\al} {\Dta_2^2 \by E_{dm}}.
\eeq
\Eno{mvst} is the experimentally observed $\sqrt{t}$ law.

As noted in \Sno{intro}, a very pretty heuristic argument for this result is
given in Ref.~\cite{gsvbook}. These authors reach the same conclusion by arguing
that $\sig(t)$ must be of the order of the typical dipole field when the spins start
flipping, and thus proportional to $a^3/\ell^3(t)$, where $\ell(t)$ is the typical
distance between reversed spins. They then note that
$a^3/\ell^3(t) \propto n_{\dn}(t)$, so that $\sig(t) \propto n_{\dn}(t)$. They then
estimate $m_r(t)$ as $W/\sig(t)$, from which it follows that $dm/dt \sim 1/n_{\dn}(t)$,
and that $n_{\dn}(t) \sim t^{1/2}$. As part of this argument, one has that
$\sig(t) \sim t^{1/2}$ and that $n_r(t) \sim t^{-1/2}$. We find the same behavior for
these quantities, but we arrive at it in a different (and more difficult!) way since we
did not have enough confidence in our understanding of the relation between $\sig(t)$ and
$n_{\dn(t)}$. Instead,
we find the delicate noncancellation between $n_r$ and ${\cal F}$ in order to first find
the differential equation obeyed by $n_r(t)$, and determine that $n_r(t) \sim t^{-1/2}$,
after which the equation for $m(t)$ is elementary. The agreement with \cite{gsvbook}
gives us encouragement that our procedure is correct, and that we can use our more
detailed rate equations to analyze other experimental protocols in the future.

\acknowledgments
This work was begun with support from the NSF via grant number
DMR-0202165. We are indebted to Rahul Pandit and Nandini Trivedi
for useful comments on Monte Carlo techniques.

\newpage
\voffset=1.0in
\hoffset=0.5in
\begin{figure}
\includegraphics{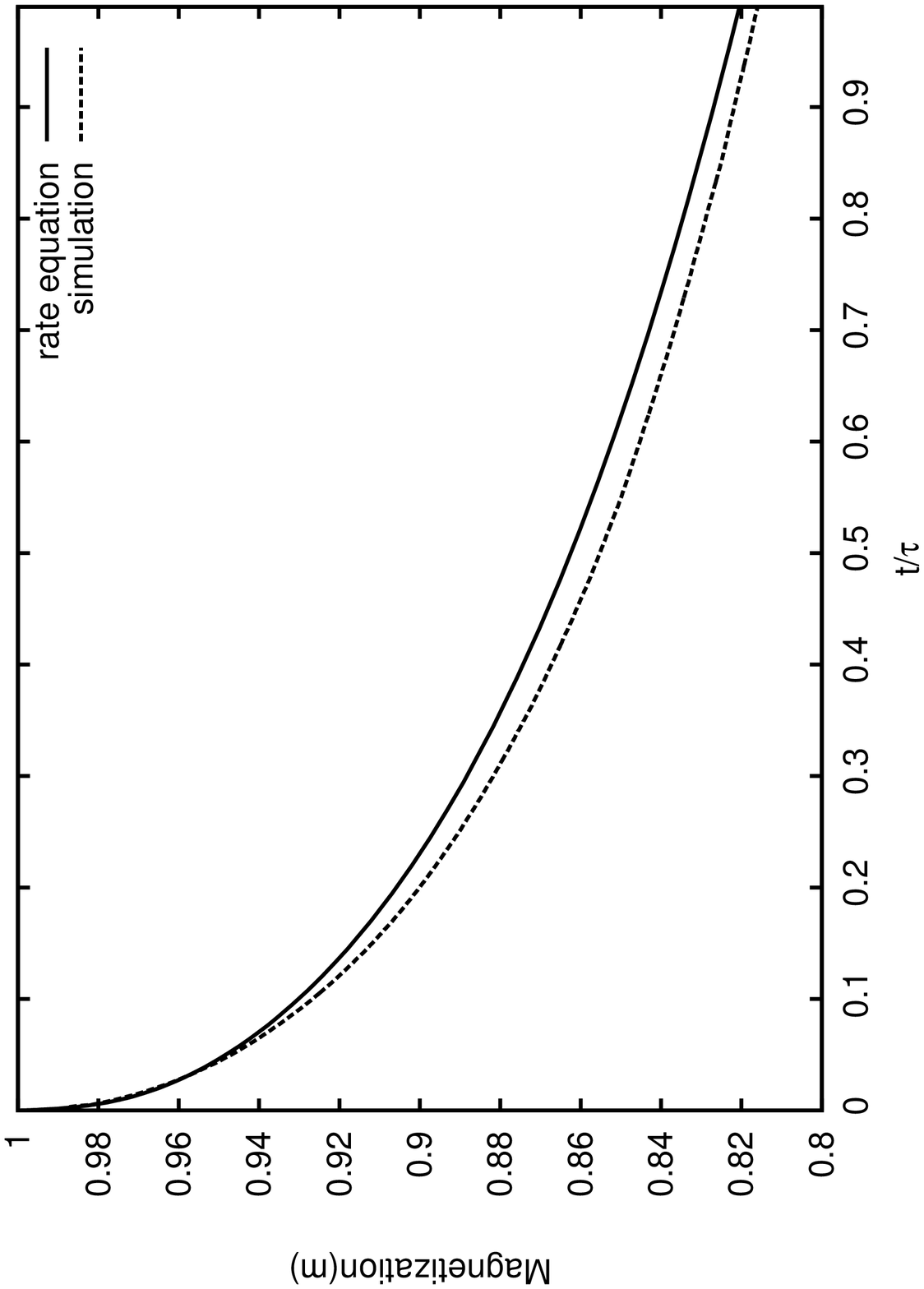}
\caption{\label{mvt_direct.ps}
Same as \fno{m_vs_t_short.eps}, except that the rate equations are solved using $\kap_2 = 50$.
} 
\end{figure}
\newpage
\voffset=-0.3in
\hoffset=-0.5in
\begin{figure}
\includegraphics{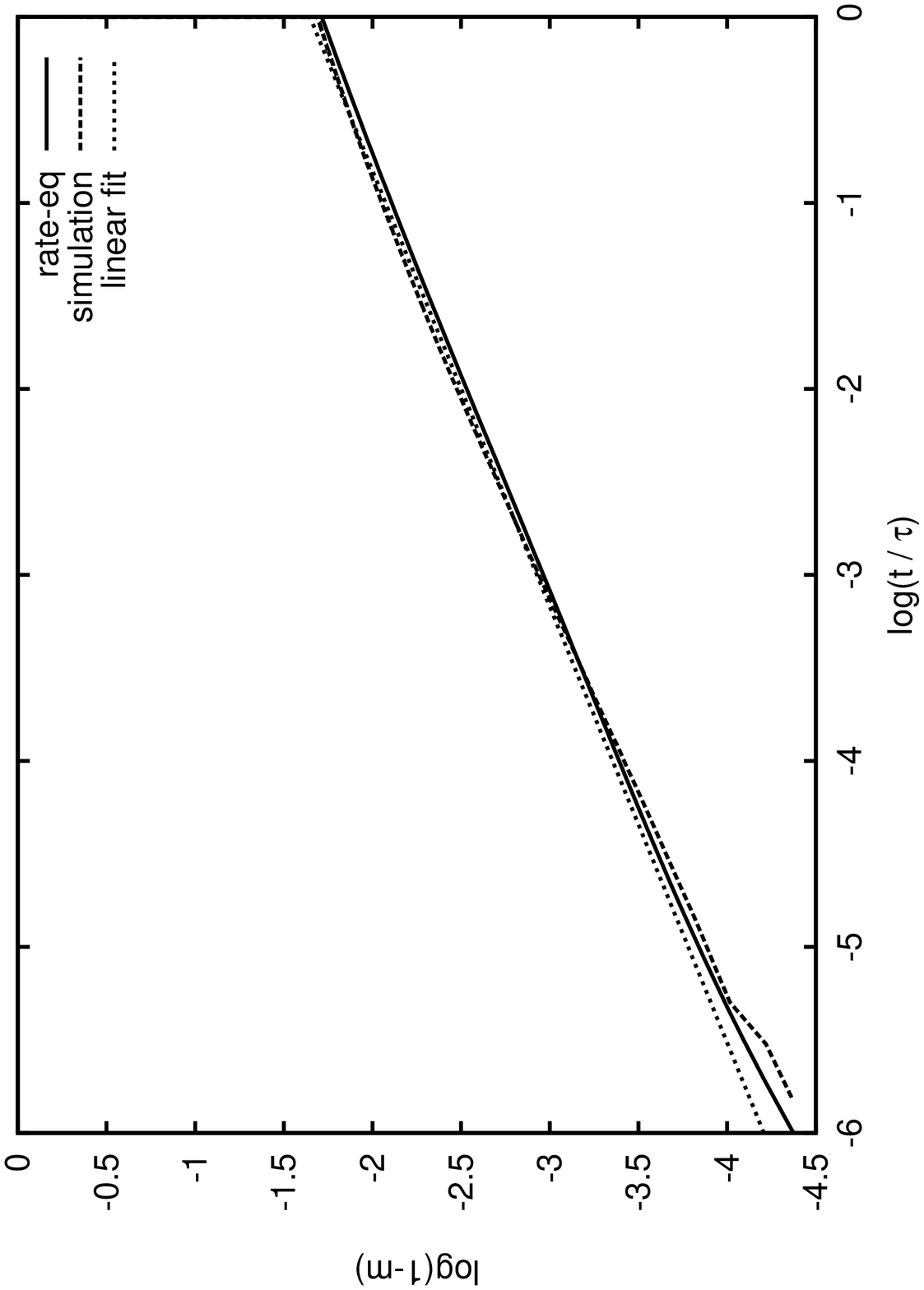}
\caption{\label{mvt_loglog.ps}
Log-log plot of the short-time behavior of the magnetization. Also shown is the solution
given by the rate equations with $\kap_2 =50$, and a linear fit to the latter. This fit gives
an exponent equal to 0.46, close to 0.5 for an exact square root.
} 
\end{figure}


\begin{thebibliography}{99}
\bibitem{bias} It is apparent that the bias $E_i = 2\mu H_i$, where
$\mu$ is the magnetic moment of a molecule, and $H_i$ is the magnetic
field at the $i$th site. We prefer to work with the bias as all
interactions in the problem are then expressed in terms of energies.
\end{thebibliography}
\end{document}